\documentclass[fleqn,usenatbib]{mnras}

\usepackage{newtxtext,newtxmath}
\usepackage[T1]{fontenc}
\usepackage{ae,aecompl}

\usepackage{graphicx}	% Including figure files
\usepackage{amsmath}	% Advanced maths commands
\usepackage{amssymb}	% Extra maths symbols

\title[WALLABY ES - I. The NGC\,7162 Galaxy Group]{WALLABY Early Science - I. The NGC\,7162 Galaxy Group.}

\author[T.N.~Reynolds, et al.]{T.N.~Reynolds$^{1,2,3}$\thanks{tristan.reynolds@research.uwa.edu.au},
T.~Westmeier$^{1,3}$,
L.~Staveley-Smith$^{1,3}$,   
A.~Elagali$^{1,2,3}$,
B.-Q.~For$^{1,3}$,
\newauthor D.~Kleiner$^{2}$,
B.S.~Koribalski$^{2,3}$,
K.~Lee-Waddell$^{2}$,
J.P.~Madrid$^{2}$,
A.~Popping$^{1,4}$,
\newauthor J.~Rhee$^{1,3}$,
M.~Whiting$^{2}$,
O.I.~Wong$^{1,3}$,
L.J.M.~Davies$^{1}$,
S.~Driver$^{1}$,
A.~Robotham$^{1,3}$,
\newauthor J.R.~Allison$^{3,5}$,
G.~Bekiaris$^{2}$,
J.D.~Collier$^{2,7,8}$,
G.~Heald$^{3,6}$,
M.~Meyer$^{1,3}$,
\newauthor A.P.~Chippendale$^{2}$,
A.~MacLeod$^{2}$,
M.A.~Voronkov$^{2}$\\
$^1$International Centre for Radio Astronomy Research (ICRAR), The University of Western Australia,\\ 35 Stirling Hwy, Crawley, WA, 6009, Australia\\
$^2$CSIRO Astronomy and Space Science, Australia Telescope National Facility, P.O. Box 76, Epping NSW 1710, Australia\\
$^3$ARC Centre of Excellence for All Sky Astrophysics in 3 Dimensions (ASTRO 3D)\\
$^4$ARC Centre of Excellence for All-sky Astrophysics (CAASTRO)\\
$^5$Sub-Dept. of Astrophysics, Department of Physics, University of Oxford, Denys Wilkinson Building, Keble Rd., Oxford, OX1 3RH, UK\\
$^6$CSIRO Astronomy and Space Sciences, Australia Telescope National Facility, PO Box 1130, Bentley, WA 6102, Australia\\
$^7$School of Computing, Engineering and Mathematics, Western Sydney University, Locked Bag 1797, Penrith, NSW 2751, Australia\\
$^8$The Inter-University Institute for Data Intensive Astronomy (IDIA), Department of Astronomy, University of Cape Town,\\ Rondebosch, 7701, South Africa}

\date{Accepted XXX. Received YYY; in original form ZZZ}

\pubyear{2018}

\begin{document}
\label{firstpage}
\pagerange{\pageref{firstpage}--\pageref{lastpage}}
\maketitle

\begin{abstract}
We present Widefield ASKAP L-band Legacy All-sky Blind Survey (WALLABY) early science results from the Australian Square Kilometre Array Pathfinder (ASKAP) observations of the NGC\,7162 galaxy group. We use archival HIPASS and Australia Telescope Compact Array (ATCA) observations of this group to validate the new ASKAP data and the data reduction pipeline ASKAP\textsc{soft}. We detect six galaxies in the neutral hydrogen (H\,\textsc{i}) 21-cm line, expanding the NGC\,7162 group membership from four to seven galaxies. Two of the new detections are also the first H\,\textsc{i} detections of the dwarf galaxies, AM\,2159-434 and GALEXASC J220338.65-431128.7, for which we have measured velocities of $cz=2558$ and $cz=2727$\,km\,s$^{-1}$, respectively. We confirm that there is extended H\,\textsc{i} emission around NGC\,7162 possibly due to past interactions in the group as indicated by the $40^{\circ}$ offset between the kinematic and morphological major axes for NGC\,7162A, and its H\,\textsc{i} richness. Taking advantage of the increased resolution (factor of $\sim1.5$) of the ASKAP data over archival ATCA observations, we fit a tilted ring model and use envelope tracing to determine the galaxies' rotation curves. Using these we estimate the dynamical masses and find, as expected, high dark matter fractions of $f_{\mathrm{DM}}\sim0.81-0.95$ for all group members. The ASKAP data are publicly available.
%This is a simple template for authors to write new MNRAS papers.
%The abstract should briefly describe the aims, methods, and main results of the paper.
%It should be a single paragraph not more than 250 words (200 words for Letters).
%No references should appear in the abstract.
\end{abstract}

\begin{keywords}
galaxies: kinematics and dynamics -- galaxies: groups  -- galaxies: distances and redshifts -- radio lines: galaxies -- instrumentation: interferometers -- telescopes
\end{keywords}

%%%%%%%%%%%%%%%%%%%%%%%%%%%%%%%%%%%%%%%%%%%%%%%%%%

%%%%%%%%%%%%%%%%% BODY OF PAPER %%%%%%%%%%%%%%%%%%

\section{Introduction} 

A galaxy's environment is known to have a strong impact on its morphology. \cite{Dressler1980} first demonstrated the morphology-density relation from optical observations, where, with increasing galaxy density, the fraction of early-type, ellipticals increases and the fraction of late-type, spirals and irregulars decreases. The morphology-density relation indicates that over-dense environments shape the evolution of galaxies through the merger history and the accretion and stripping of neutral hydrogen gas, H\,\textsc{i}, which affects star-formation history. Radio observations of H\,\textsc{i} provide another window on the environmental impact in situ. H\,\textsc{i} observations show a dependence of gas content with environment as spiral galaxies in dense environments, such as towards the centre of clusters, often have less H\,\textsc{i} and reduced star formation rates compared to isolated, field galaxies of the same size and morphology \citep[e.g.][]{Giovanelli1985,Solanes2001,Denes2014,Odekon2016}. The H\,\textsc{i} mass function has also been shown to vary with galaxy environment and morphological type \citep[e.g.][]{Zwaan2005,Jones2018}. It follows that the density of the group and cluster environments influence the physical mechanisms responsible for changing galaxy H\,\textsc{i} composition and morphology. 

Ram pressure stripping \citep[the removal of H\,\textsc{i} gas and stars from a galaxy passing through a dense intergalactic medium,][]{Gunn1972,Chung2009} and harassment \citep[interacting galaxies with high relative velocities,][]{Moore1996,Moore1998} are more common in clusters. In small groups, tidal stripping \citep[low relative velocities between interacting galaxies,][]{Moore1999,Koribalski2009,English2010} and starvation \citep[hot gas removed from the galaxy's extended halo, disrupting gas accretion, cutting off further gas infall onto the galaxy and quenching of star formation,][]{Larson1980} are more typical. Physical mechanisms more commonly seen in clusters also act in groups. For instance, ram pressure stripping has been observed even in low density groups \citep[e.g.][]{Rasmussen2006,Rasmussen2012,Westmeier2011}.

The H\,\textsc{i} content and physical mechanisms affecting galaxies in high density cluster environments have been well studied in the local Universe \citep[e.g.][]{Kenney2004,Jaffe2015}, in part due to the ability to efficiently observe a large number of galaxies in a small area of sky. Lower density group environments have been less studied due to the significantly increased telescope time required to obtain a galaxy sample comparable to even a single cluster. 

Groups are important as they are the most common environment in which to find galaxies \citep[e.g.][]{Tully1987,Gourgoulhon1992}. The environment begins to affect the H\,\textsc{i} content and evolution of galaxies while they are in low density groups. This physical process is known as pre-processing \citep[e.g.][]{Wevers1984,Zabludoff1998,Kern2008,Freeland2009,Kilborn2009,Koribalski2012,Hess2013}. In recent years, morphological and kinematic studies of individually resolved galaxy groups \citep[e.g.][]{Koribalski2004,Koribalski2005,Kilborn2005,Serra2013,Serra2015b,Hess2017} and surveys of a few dozen resolved groups \citep[e.g.][]{Brough2006,Kilborn2009,Pisano2011} have begun to build up a picture of H\,\textsc{i} in the group environment. However, large statistical samples currently only exist of global H\,\textsc{i} properties from surveys on single dish telescopes \citep[e.g. HIPASS and ALFALFA,][respectively]{Barnes2001,Haynes2018}. Complementary to group studies are deep interferometric surveys of individual galaxies, such as the Westerbork Hydrogen Accretion in LOcal GAlaxieS survey \citep[HALOGAS,][]{Heald2011} and the ongoing Imaging Galaxies Inter-galactic and Nearby Environment survey (IMAGINE, PI A. Popping) and the MeerKAT Observations of Nearby Galactic Objects - Observing Southern Emitters survey \citep[MHONGOOSE,][]{deblok2017} providing a comparison to the group environment. Large surveys of resolved isolated galaxies and galaxy groups require the ability to replicate the survey speed of single dish telescopes, but with increased resolution only achievable with an interferometer.

Additionally, resolved H\,\textsc{i} observations are used for estimating the total dynamical and dark matter masses of galaxies using their rotation curves \citep[e.g.][]{deblok2001,deblok2008,Sofue2001,Oh2008,Oh2011,Oh2015,Westmeier2011,Westmeier2013}. This is also one of the goals of the Local Volume H\,\textsc{i} Survey \citep{Koribalski2018,Oh2018}. A clear advantage of interferometric H\,\textsc{i} surveys, like WALLABY, will be the ability to map the dark matter distribution in gas-rich galaxies across the entire southern sky. If the central $\sim2$\,kpc are resolved, H\,\textsc{i} rotation curves can also be used to differentiate among various proposed dark matter density profiles \citep[e.g. pseudo-isothermal and NFW,][respectively]{Begeman1991,Navarro1997}. 

% ----------------------------------------------------------------------------------------------------
\subsection{ASKAP}
\label{s-sec:askap_intro}

Traditionally, radio telescopes built from paraboloidal reflector antennas have used large, single pixel feed-horn receivers, which limits the number of receivers that can be placed simultaneously at the focus of the antenna (e.g. 13 on the Parkes radio telescope multi-beam receiver). Phased array feeds (PAFs) are a recently developed type of receiver consisting of a plane of antenna elements that can form multiple beams on the sky simultaneously. The CSIRO Australian Square Kilometre Array Pathfinder (ASKAP) telescope is fitted with PAFs \citep{DeBoer2009,Hampson2012,Hotan2014,Schinckel2016}, which consist of 188 connected dipoles in a chequerboard pattern \citep{Hay2008} and can form up to 36 dual-polarisation beams on the sky, simultaneously covering a significantly larger area in a single pointing than traditional receivers. PAFs are the ideal receiver to expand an interferometer's instantaneous field of view, as antennas in an interferometric array are generally too small to accommodate multiple feed-horn receivers at the focus (i.e. 12\,m vs 64\,m diameter dish for ASKAP and Parkes, respectively). 

The Widefield ASKAP L-band Legacy All-sky Blind Survey \citep[WALLABY,][]{Koribalski2012} is one of the surveys that will take advantage of the wide field of view of the ASKAP PAFs. WALLABY will cover $\sim75\%$ of the sky observing an estimated 500000 galaxies in  H\,\textsc{i} out to $z<0.26$ \citep{Duffy2012}. Prior to the survey commencing, several early science fields, each a single 30\,deg$^2$ field using 12 antennas (ASKAP-12) and limited bandwidth, have been observed for testing of ASKAP and validation of the ASKAP data reduction pipeline (ASKAP\textsc{soft}). This work presents early science observations carried out with reduced bandwidths of 48, 192 and 240\,MHz rather than the full ASKAP bandwidth of 304\,MHz.

% and forms one of a number of WALLABY early science papers.

% ----------------------------------------------------------------------------------------------------
\subsection{The NGC 7162 Galaxy Group}
\label{s-sec:group_background}

In this work we present the first ASKAP observations of the galaxy group consisting of NGC\,7162, NGC\,7162A, NGC\,7166 and ESO\,288-G025 \citep{Maia1989,Fouque1992}. We also detect three additional galaxies, which we identify as possible group members: ESO\,288-G033, AM\,2159-434 and GALEXASC J220338.65-431128.7. The galaxies cover a velocity range of $\sim2150-2750$\,$\mathrm{km}\,\mathrm{s}^{-1}$ and are located within a $\sim1.5\times1.5$ square degree area centred on $\alpha,\delta=$ 22:01:00,$-$43:30:00; J2000. If we assume a distance to the group of $\sim33.7$\,Mpc, NGC\,7162 and NGC\,7162A have a projected separation of $\sim140$\,kpc and the most distant group member, ESO\,288-G025, has a projected separation of $\sim332$\,kpc from NGC\,7162. We list archival and new galaxy parameters in Table~\ref{table:galaxy_params}. For computed galaxy luminosity distances, we assume velocity uncertainties of 200\,$\mathrm{km}\,\mathrm{s}^{-1}$ from peculiar velocities \citep[e.g.][]{Springob2014}, giving distance uncertainties of $\sim3.0$\,Mpc. The NGC\,7162 group is $\sim3.8$\,degrees to the north-west of the NGC\,7232 triplet at the centre of the WALLABY early science field. Results of NGC\,7232 triplet and IC\,5201, also located near the centre of the NGC\,7232 field, will be presented in Lee-Waddell et al. (in prep.) and Kleiner et al. (in prep.), respectively. There are also archival observations covering NGC\,7162, NGC\,7162A and ESO\,288-G033 taken with the Australia Telescope Compact Array (ATCA), which we use for validation of the ASKAP data. 

We briefly summarise earlier radio observations of the different galaxies of the NGC\,7162 group. NGC\,7162 and NGC\,7162A are late type spirals, detected and included in the H\,\textsc{i} Parkes All-Sky Survey catalogue \citep[HIPASS, sources HIPASS J2159--43 and HIPASS J2200--43, respectively,][]{Meyer2004}. Note however, the HIPASS detection of NGC\,7162 is confused with NGC\,7162A due to the Parkes beam size ($\sim15.5$\,arcmin), so the HIPASS measured spectrum and total flux for both galaxies are not separated. ESO\,288-G025 and ESO\,288-G033 are late type spirals, which are marginally detected in the HIPASS data, but not included in the HIPASS catalogue. AM\,2159-434 and GALEXASC J220338.65-431128.7 are dwarf galaxies with no previous H\,\textsc{i} detection or distance measurement.

This paper is structured as follows. In Section~\ref{sec:data} we describe the ASKAP and ATCA observations and data reduction process. We present H\,\textsc{i} moment maps and spectra in Section~\ref{sec:analysis} and our validation of the ASKAP observations and processing pipeline (ASKAP\textsc{soft}\footnote{Complete documentation of the current ASKAP\textsc{soft} version can be found at \url{http://www.atnf.csiro.au/computing/software/askapsoft/sdp/docs/current/index.html}}). We use tilted-ring modelling and envelope tracing to derive the rotation curves and carry out mass modelling to estimate the dark matter mass in the galaxies in Sections~\ref{sec:rotcur} and \ref{sec:mass_mod}. In Section~\ref{sec:hi_gas} we calculate the H\,\textsc{i} gas mass and deficiencies of the group spiral galaxies. We present our discussion and conclusions in Sections~\ref{sec:discussion} and \ref{sec:conclusion}, respectively. 

Throughout, we use J2000 coordinates, dates in UTC, velocities in the optical convention ($cz$) and the heliocentric reference frame. Galaxy quantities are calculated using distances derived from velocities converted to the local group (LG) frame\footnote{Heliocentric to local group reference frame velocity correction calculated using the NED Velocity Correction Calculator, \url{https://ned.ipac.caltech.edu/forms/vel_correction.html}}, adopting a flat $\Lambda$CDM cosmology using ($H_0$, $\Omega_{\mathrm m}$) $=$ (67.7, 0.307), concordant with the latest \textit{Planck} results \citep{Planck2016}.

% -------------------------------------------------------------------------------------------------------------------------------------------------------------------------
\section{OBSERVATIONS AND DATA REDUCTION}
\label{sec:data}

% ----------------------------------------------------------------------------------------------------
\subsection{ASKAP}
\label{s-sec:askap_obs}

NGC\,7162A and NGC\,7162 lie in the top right corner (north west) beams of the WALLABY early science field centred on the NGC\,7232 galaxy group ($\sim2.5\times2.5$ square degrees centred on $\alpha,\delta=$ 22:00:00.0,$-$43:30:00.0). WALLABY early science observations are carried out with beams arranged in a $6\times6$ square grid resulting in a field of view of 30 square degrees. This beam arrangement is used for two footprints on the sky (A, centred on $\alpha,\delta=$ 22:13:07.7,$-$45:16:57.1, and B, centred on $\alpha,\delta=$ 22:10:35.41,$-$44:49:50.7), with footprint B offset by 0\fdg64 from footprint A. This configuration allows us to obtain uniform sensitivity across the sky, within the overlap region, by combining the two footprints. ASKAP beams are formed by pointing the antennas at the Sun prior to the start of observing using the maximum signal- to-noise ratio \citep[maxSNR,][]{Applebaum1976} method \citep[for details see][]{Chippendale2015,McConnell2016}. This early science field was observed over 14 nights in August and October of 2016 and 2 nights in August and September of 2017 for a total of 175.3\,hrs with 8 nights in footprint A and 8 nights in footprint B (see Fig.~\ref{fig:ngc7162_field} for the beam positions on the sky for footprint A and B, green and magenta circles, respectively). These observations used an array of $10-12$ antennas from ASKAP-12. We use only 6 beams (2 footprint A, 4 footprint B) and a limited bandwidth of 8\,MHz (432 channels) centred on the group (1409.56\,MHz) to keep data volume and computing requirements manageable, in order to facilitate software testing and debugging. Table~\ref{table:obs_params} summarises the observations from both ASKAP and ATCA.

% --------------------------------------------------------------------
\subsubsection{ASKAPSOFT}
\label{ss-sec:askapsoft}

We reduced the ASKAP observations using a preliminary version of ASKAP\textsc{soft}, the data reduction pipeline built for handling ASKAP data. Full details on ASKAP\textsc{soft} will be presented in Whiting et al. (in prep.) and Kleiner et al. (in prep.). We only give a brief summary here as our procedure only deviates at a couple steps from the standard pipeline. ASKAP\textsc{soft} first splits visibilities formed for individual beams from the observation. These are then flagged, have bandpass and gain calibrations applied and are imaged separately before mosaicking beam images together. Our primary calibrator is PKS\,1934-638, which we use for flux and bandpass calibration. Some additional manual flagging of the first two hours from the footprint B visibilities was required on the shortest baseline to remove solar interference. After flagging and calibrating the observations, ASKAP\textsc{soft} creates a continuum sky model which is used to perform continuum subtraction on the spectral uv data (i.e. the model is used to simulate visibilities which are then subtracted from the observed visibilities).

The default ASKAP\textsc{soft} pipeline can only handle a single night's observation for imaging, limiting the depth to which we can \textsc{clean} to three times the root-mean-square (RMS) noise level of a single night ($3\sigma\sim18$\,mJy\,beam$^{-1}$). Image combination is then carried out in the image plane by linear mosaicking the beams from individual nights. To lower the threshold to which we can \textsc{clean}, we used a modified pipeline script for imaging to combine data in the uv-domain. Using this script, we can feed the imager multiple nights of uv-domain data for a single beam and footprint. We imaged using a resolution of 5\arcsec\,pixel$^{-1}$ and 4\,$\mathrm{km}\,\mathrm{s}^{-1}$ channels, with \textsc{robust} $=0.5$ and a Gaussian taper of 30\arcsec, resulting in a synthesised beam of $39\arcsec\times34\arcsec$. With the modified imaging script we imaged 8 nights worth of data, lowering the RMS and improving our deconvolution ($3\sigma\sim9$\,mJy\,beam$^{-1}$). We do not achieve a $\sqrt{8}$ improvement in the RMS due to different integration times, percentage of flagged data and number of antennas used each night (see Section~\ref{s-sec:validation}). For deconvolution, we use the multi-scale \textsc{clean} algorithm \citep[e.g.][]{Cornwell2008a,Rau2011} on scales of 0 (point sources), 3, 10 and 30\,pixels. We set the major cycle \textsc{clean} threshold to 3\,$\sigma$ and for the minor cycle 4.5\,$\sigma$. These parameters were fine-tuned to maximize the amount of flux recovered after deconvolution. Prior to mosaicking the imaged beams, we first removed residual continuum emission, visible due to the decreased RMS level ($\sim3$\,mJy\,beam$^{-1}$), using image-based continuum subtraction (e.g. subtracting a 2$^{\mathrm{nd}}$ order polynomial fit to residual continuum flux from the data cube). Our final mosaicked cube RMS level is $\sim2.3$\,mJy\,beam$^{-1}$. We note that full ASKAP (36 antennas) will be much more sensitive than ASKAP-12 and will reach WALLABY sensitivity (1.6\,mJy\,beam$^{-1}$) in a single 12\,hr observation.

% --------------------------------------------------------------------
\subsubsection{MIRIAD}
\label{ss-sec:askap_miriad}

In addition to performing imaging with ASKAP\textsc{soft}, we also imaged using \textsc{miriad} \citep{Sault1995} for validation purposes. Similar to ASKAP\textsc{soft}, we imaged all the flagged, calibrated and continuum subtracted uv-data beam by beam with the task \textsc{invert}, using the same pixel and spectral resolution, robustness and taper from ASKAP\textsc{soft}. We deconvolved the dirty image using the \textsc{clean} algorithm with a cutoff flux of 5\,mJy and 8000 iterations and restored the deconvolved image using \textsc{restor}. We used the task \textsc{contsub} to fit and subtract a 0$^{\mathrm{th}}$ order polynomial to the emission free channels to remove residual continuum emission, similar to ASKAP\textsc{soft}. We finally create a mosaicked image cube using the task \textsc{linmos}, applying primary beam correction assuming a $1\degr\times1\degr$ Gaussian and weighting each beam cube by the image RMS. The RMS of the mosaicked \textsc{miriad} imaged data is comparable to that obtained imaging with ASKAP\textsc{soft} ($\sim$2.2 vs 2.3\,mJy\,beam$^{-1}$, respectively). The synthesised beam, $37\arcsec\times32\arcsec$, is $\sim2\arcsec$ smaller than the one produced using ASKAP\textsc{soft}. We note that \textsc{miriad} is unable to image the ASKAP data perfectly as it does not account for non-coplanar baselines, which can be accounted for using the w-projection algorithm \citep{Cornwell2008b}, and may cause position shifts and minor imaging artefacts.

% ----------------------------------------------------------------------------------------------------
\subsection{ATCA}
\label{s-sec:atca_obs}

\begin{figure*}
	\centering
	\includegraphics[width=13cm]{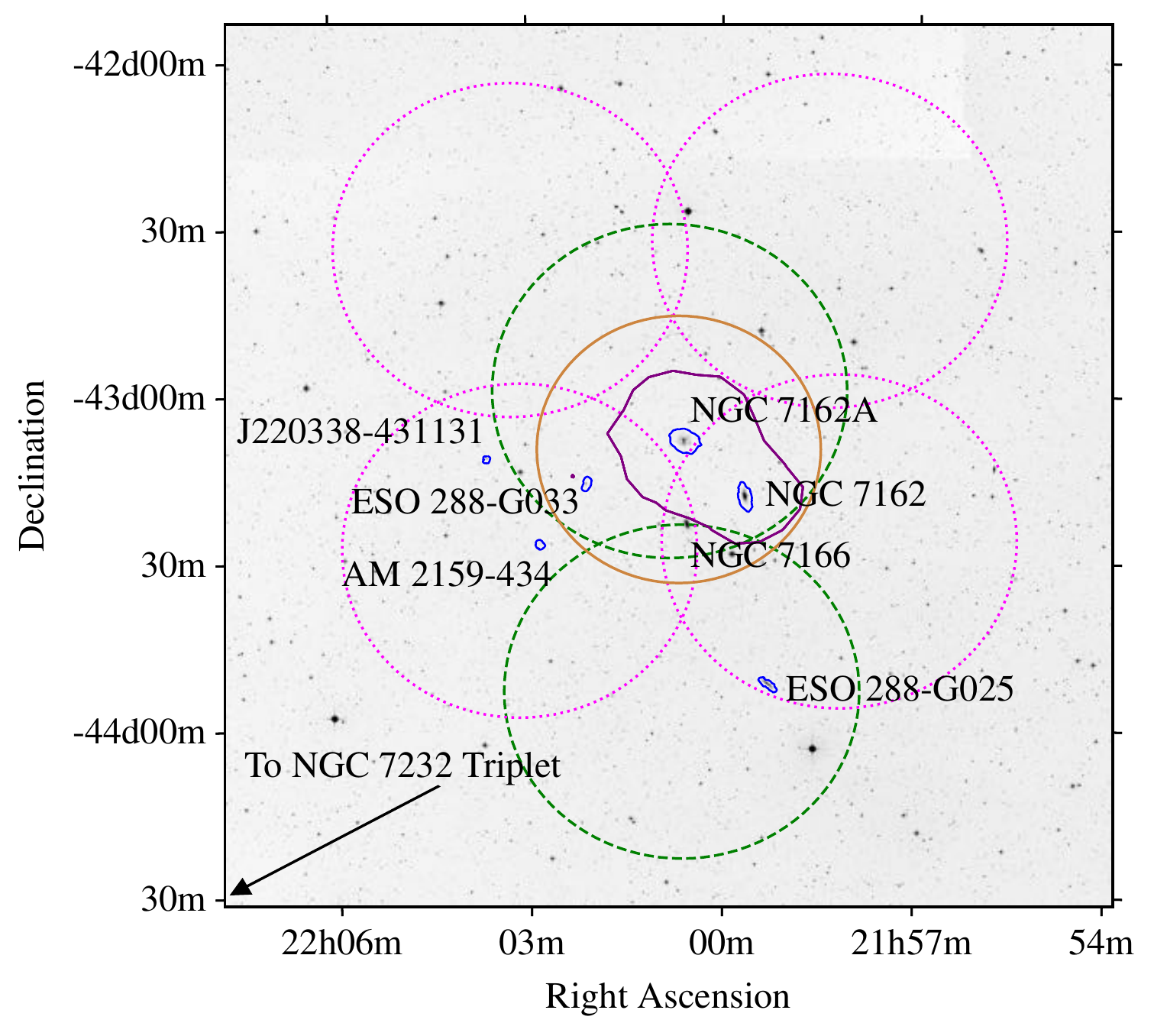}
		\caption{Digitized Sky Survey blue optical image with $1.6\times10^{19}\,\mathrm{cm}^{-2}$ and $2.8\times10^{18}\,\mathrm{cm}^{-2}$ H\,\textsc{i} column density contours overlaid from ASKAP (blue) and HIPASS (purple). The observed footprint is shown for the ASKAP footprints A (green, dashed circles) and B (magenta, dotted circles) and ATCA (orange, solid circle). The footprint A beams are numbers 16 and 35 (top and bottom, respectively). The footprint B beams are numbers 16, 35, 04, 17 (clockwise from top right). Beam numbers are from the full 36 beam footprint. The circles indicate the nominal full width at half maximum (FWHM) for each beam.}
	\label{fig:ngc7162_field}
\end{figure*}

The archival Australia Telescope Compact Array (ATCA) observations of NGC\,7162, NGC\,7162A and ESO\,288-G033, were obtained under project ID C2573 (Observer: S. Reeves) for a single pointing centred on NGC\,7162A (Fig.~\ref{fig:ngc7162_field}, solid orange circle), using the 750D antenna configuration \citep{Reeves2015} and the Compact Array Broadband Backend system \citep[CABB,][]{Wilson2011}. The flux and phase calibrators observed were PKS\,1934-638 and PKS\,2106-413, respectively. The total on-source integration time was 994\,min (see Table~\ref{table:obs_params}).

We reduced the data using \textsc{miriad} using a standard method. After first excluding antenna 6, we interactively flagged the flux and phase calibrators and science field data using the task \textsc{blflag}. We then calibrated the flux of PKS\,1934-638 observations using \textsc{mfcal}. We applied the flux calibration to the phase calibrator, PKS\,2106-413, and determined time-dependent gain solutions on the phase calibrator. We then applied both phase and gain calibrations to the science observations.

We inspected the time-integrated shortest baseline (antenna 1-2) to find channels clear of any H\,\textsc{i} emission to find the continuum, which we subtracted using \textsc{uvlin} with a 2$^{\mathrm{nd}}$ order polynomial. We created a dirty map from the continuum subtracted data setting the weighting to \textsc{robust} $=0.5$ and a channel width of 10\,km\,s$^{-1}$. We use a spectral resolution of 10\,km\,s$^{-1}$ instead of the raw resolution of 6.65\,km\,s$^{-1}$ to improve our signal to noise. We then deconvolved the dirty map using H\"ogbom \textsc{clean} with a 3\,$\sigma$ cutoff flux of 6\,mJy\,beam$^{-1}$ and 10000 iterations before restoring the deconvolved image cube with \textsc{restor}. The resulting data cube has an RMS of 1.2\,mJy\,beam$^{-1}$, which agrees with the expected RMS for ATCA with these observation parameters (1.1\,mJy\,beam$^{-1}$), and a synthesised beam of $60\arcsec\times36\arcsec$.

\begin{table*}
	\centering
    \caption{ATCA and ASKAP observation parameters.}
	\label{table:obs_params}
	\begin{tabular}{lcccr}
		\hline
		Observation & ATCA & ASKAP & ASKAP & ASKAP \\
		Parameter & & 48\,MHz & 192\,MHz & 240\,MHz  \\ \hline
		Project ID & C2573 & AS035 & AS035 & AS035 \\
		Dates & 2013 Aug 2--3 & 2016 Aug 11--12 & 2017 Aug 23 & 2017 Sept 27 \\
		 &  & 2016 Oct 7--19 & & \\
		Configuration & 750D & ASKAP-12 & ASKAP-12 & ASKAP-12 \\
        Minimum Baseline & 31\,m & $22-61$\,m & $22-61$\,m & $22-61$\,m \\
        Maximum Baseline & 719\,m & 2300\,m & 2300\,m & 2300\,m \\
		Integration Time & 994\,min & 160.3\,hr & 10\,hr & 5\,hr \\
		Bandwidth & 64\,MHz & 48\,MHz & 192\,MHz & 240\,MHz \\
		Channels & 2048 & 2592  & 10368 & 12960 \\
		Channel Width & 31.25\,kHz & 18.5\,kHz & 18.5\,kHz & 18.5\,kHz \\
		Central Frequency & 1406.0\,MHz & 1400.497\,MHz & 1344.5\,MHz & 1320.5\,MHz \\
		Polarisations & XX, YY & XX, YY & XX, YY & XX, YY \\ \hline
	\end{tabular}
\end{table*}

% -------------------------------------------------------------------------------------------------------------------------------------------------------------------------
\section{IMAGE ANALYSIS}
\label{sec:analysis}

We use the Source Finding Application \citep[SoFiA,][]{Serra2015a} to locate significant H\,\textsc{i} emission in the ASKAP data cube. SoFiA provides integrated intensity (moment 0) and velocity field (moment 1) maps, integrated spectra and detected source properties, including integrated flux and line widths ($w_{20}$ and $w_{50}$). In SoFiA, we set a 5\,$\sigma$ threshold and allow the initial SoFiA source mask to expand and encompass additional voxels until the total source flux stopped increasing.

\begin{table}
%\footnotesize
	\centering
    \caption{WALLABY source name IDs and common galaxy names, detected in H\,\textsc{i}.}
	\label{table:askap_ids}
	\begin{tabular}{lr}
		\hline
		WALLABY Source ID & Galaxy \\ \hline
		WALLABY J215939$-$431822 & NGC\,7162 \\
        WALLABY J220034$-$430822 & NGC\,7162A \\
        WALLABY J215917$-$435201 & ESO\,288-G025 \\
        WALLABY J220206$-$431603 & ESO\,288-G033 \\
        WALLABY J220249$-$432652 & AM\,2159-434 \\ 
        WALLABY J220338$-$431131 & GALEXASC\,J220338.65-431128.7 \\ \hline
	\end{tabular}
\end{table}

\begin{figure*}
	\centering
	\includegraphics[width=16cm]{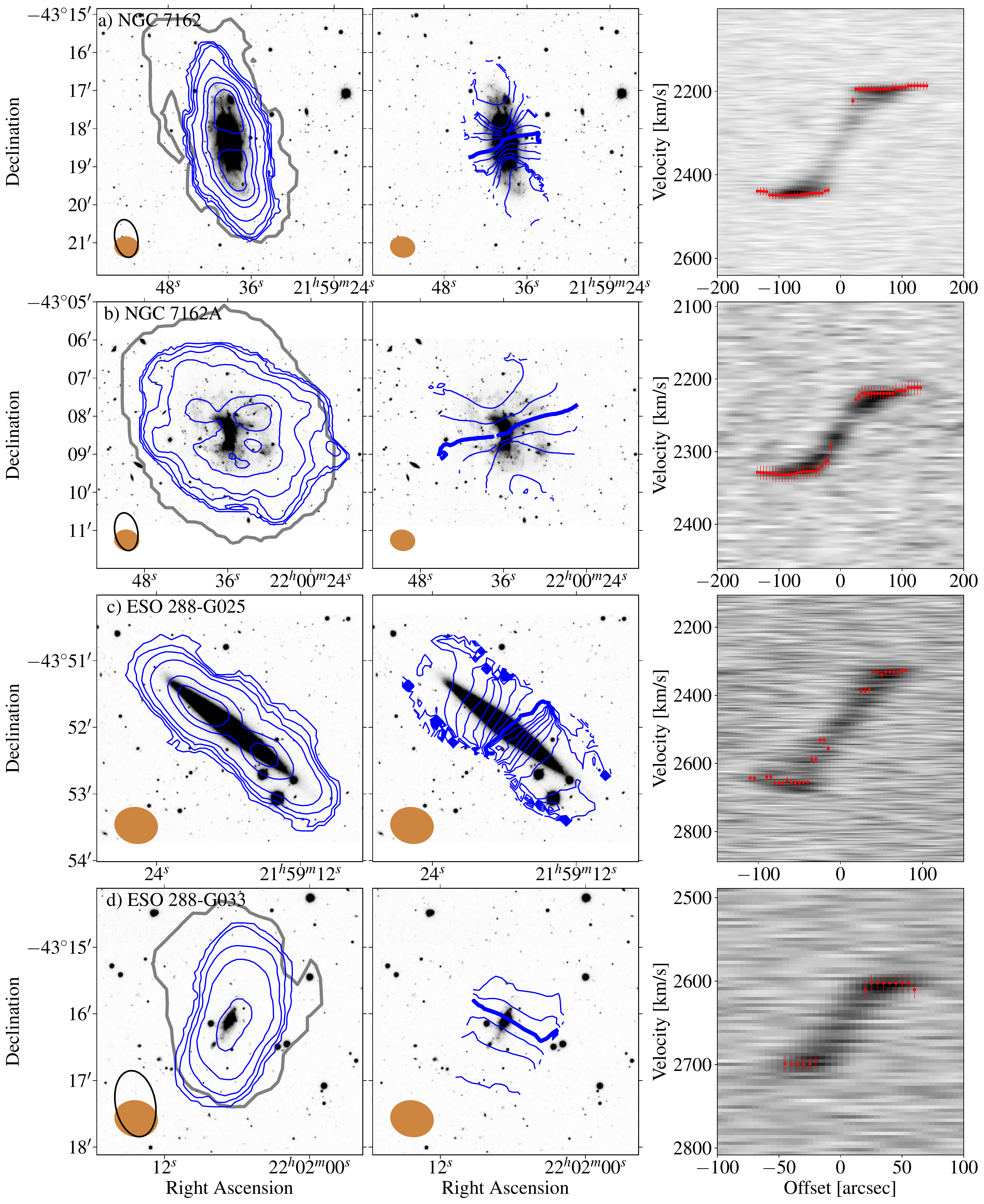}
		\caption{ASKAP H\,\textsc{i} moment maps (column density and velocity field, left hand and centre columns, respectively) of the four spiral galaxies (blue contours) overlaid onto Dark Energy Survey (DES) $r-$band grey scale images \protect\citep{Abbott2018}. The panel sizes for the column density and velocity field maps are $7\arcmin\times7\arcmin$ (a and b) and $4\arcmin\times4\arcmin$ (c and d). Column density map contour levels are (1, 5, 10, 20, 50, 70, 100)$\,\times\,10^{19}\,\mathrm{cm}^{-2}$. Velocity field (heliocentric reference frame) map contours levels in 20\,km\,s$^{-1}$ steps decreasing and increasing from the systemic velocity (thick line) of each galaxy (see Table~\ref{table:galaxy_params}). The velocity increases from the southern (lower) side (panels a and b) and increases from the northern (upper) side (panels c and d). The ASKAP synthesised beam sizes is shown in the lower left corner of the panels in the left and centre columns by the orange ellipse. We also include the $10^{19}\,\mathrm{cm}^{-2}$ H\,\textsc{i} column density contour from ATCA (thick grey contour) for NGC\,7162, NGC\,7162A and ESO\,288-G033, along with the ATCA synthesised beam (black ellipse). Position-Velocity diagrams (right hand column) are shown with terminal velocities measured using the envelope tracing method (red circles). We exclude points with an offset of $<\pm20$\arcsec from the envelope tracing.}
	\label{fig:all_maps4}
\end{figure*}

\begin{figure*}
	\centering
	\includegraphics[width=16cm]{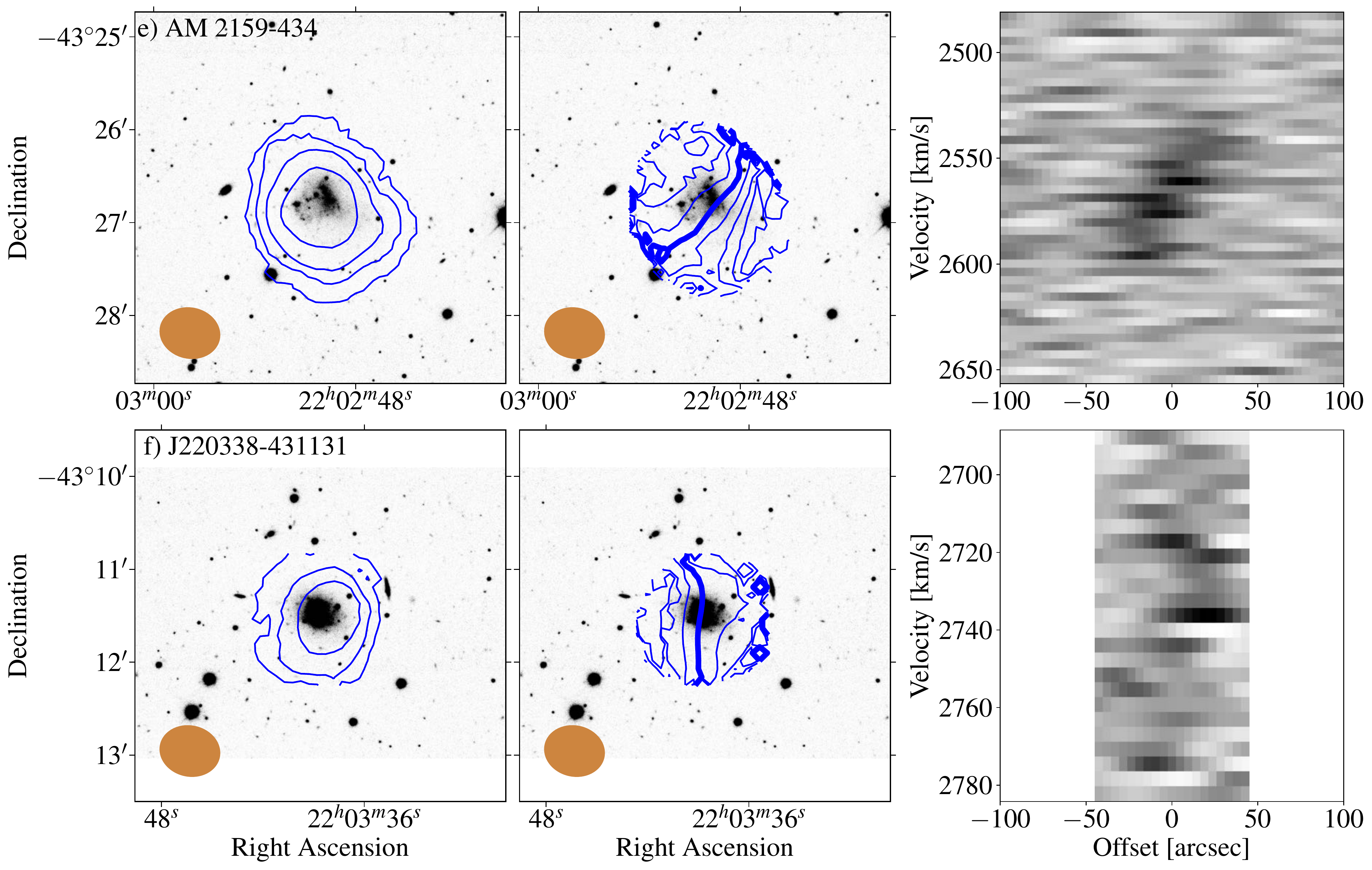}
		\caption{Similar to Fig.~\ref{fig:all_maps4} for the two dwarf galaxies with background DES $g-$band grey scale images. The panel sizes for the column density and velocity field maps are $4\arcmin\times4\arcmin$ (e and f). Column density map contour levels are (1, 5, 10, 20, 50, 70, 100)$\,\times\,10^{19}\,\mathrm{cm}^{-2}$. Velocity field (heliocentric reference frame) map contours levels in 7\,km\,s$^{-1}$ steps decreasing and increasing from the systemic velocity (thick line) of each galaxy (see Table~\ref{table:galaxy_params}). The velocity increases from the western (right) side.}
	\label{fig:all_maps2}
\end{figure*}

% ----------------------------------------------------------------------------------------------------
\subsection{HI Maps}
\label{s-sec:maps}

In our survey volume (Fig.~\ref{fig:ngc7162_field}) we detect six ASKAP H\,\textsc{i} sources using SoFiA, each identified with a known optical counterpart. The sources are named using the centre position defined by SoFiA (Table~\ref{table:askap_ids}), hereafter referred to by their more common names except for GALEXASC J220338.65-431128.7 which we refer to as `J220338-431131'. We detect the original group members, NGC\,7162, NGC\,7162A and ESO\,288-G025, while we do not detect H\,\textsc{i} in NGC\,7166, in agreement with previous ATCA observations \citep{Oosterloo2007}. In addition to the group galaxy detections, we also detect H\,\textsc{i} in two nearby dwarf galaxies, AM\,2159-434 ($\alpha,\delta=$ 22:02:50,$-$43:26:44) and J220338-431131 ($\alpha,\delta=$ 22:03:38,$-$43:11:28), with no previous H\,\textsc{i} detections and beyond the field of view of ATCA. Their projected distances from NGC\,7162A, assuming a distance of 33.7\,Mpc, are 300 and 328\,kpc, respectively. This demonstrates the power of the wide field of view of the ASKAP for discovering previously undetected dwarf galaxies. Both dwarfs are also detected in GALEX NUV/FUV imaging indicating recent star formation. We also detect ESO\,288-G033 ($\alpha,\delta=$ 22:02:06,-43:16:07), which is within the ATCA beam of the archival observations providing an additional galaxy for validation. For the three galaxies in the ATCA beam, ATCA detects H\,\textsc{i} to larger radii than ASKAP-12 as expected for the larger ATCA synthesised beam, more compact baselines and the lower H\,\textsc{i} column density sensitivity threshold used for source finding with SoFiA ($3\,\sigma=1.7\times10^{18}$\,cm$^{-2}$ and $5\,\sigma=9.5\times10^{18}$\,cm$^{-2}$ for \textsc{miriad} and ASKAP\textsc{soft}, respectively). 

In Fig.~\ref{fig:ngc7162_field}, we show a Digitized Sky Survey (DSS) blue optical image of the full group with $1.6\times10^{19}\,\mathrm{cm}^{-2}$ H\,\textsc{i} column density contours of the detected galaxies overlaid in blue, the footprint of the ASKAP beams covering the group (green and magenta circles) and indicate the direction of the NGC\,7232 triplet, located at the centre of the ASKAP 36 beam footprint. We show Dark Energy Survey (DES) $r-$ and $g-$band \citep{Abbott2018} postage stamps of each H\,\textsc{i} detected spiral ($r$) and dwarf ($g$) galaxy overlaid with integrated intensity and velocity field contours and position-velocity (PV) diagrams taken along the major axis of each galaxy from SoFiA (Fig.~\ref{fig:all_maps4} and \ref{fig:all_maps2}: left, centre and right columns, respectively).

We use a 5\,$\sigma$ threshold and mask dilation in SoFiA to avoid picking up sidelobe emission during source finding. We initially used a 3\,$\sigma$ threshold for source finding to pick up any faint emission slightly above the noise level. However, this also picks up and includes sidelobe emission around our two brightest galaxies, NGC\,7162 and NGC\,7162A, in these galaxies' source masks. The sidelobe artefacts are due to the incomplete uv-coverage of ASKAP-12 and systematic errors in the calibration of the data. SoFiA does not pick up sidelobes around our other four detections using the 3\,$\sigma$ threshold as they are fainter and any sidelobes are below the image cube noise level. The \textsc{clean} artefacts limit our ability to comment on the presence or lack of faint extended emission, as any extended emission is lost in the sidelobes using the 3\,$\sigma$ threshold and the 5\,$\sigma$ threshold will miss any faint emission. Even using a 5\,$\sigma$ threshold, SoFiA still picks up the first negative sidelobe in some channels which will tend to lower the integrated fluxes. Full ASKAP will not have the same challenges as ASKAP-12 as it will have significantly improved uv-coverage with 36 antennas and a finalised pipeline using the optimal calibration and imaging parameters. Unsurprisingly, the deconvolution is not improved using \textsc{miriad}, which has the additional w-projection issues, contributing to the flux loss in the integrated spectra (Fig.~\ref{fig:all_spec}).

% ----------------------------------------------------------------------------------------------------
\subsection{HI Spectra}
\label{s-sec:spectra}

\begin{figure}
	\centering
	\includegraphics[width=\columnwidth]{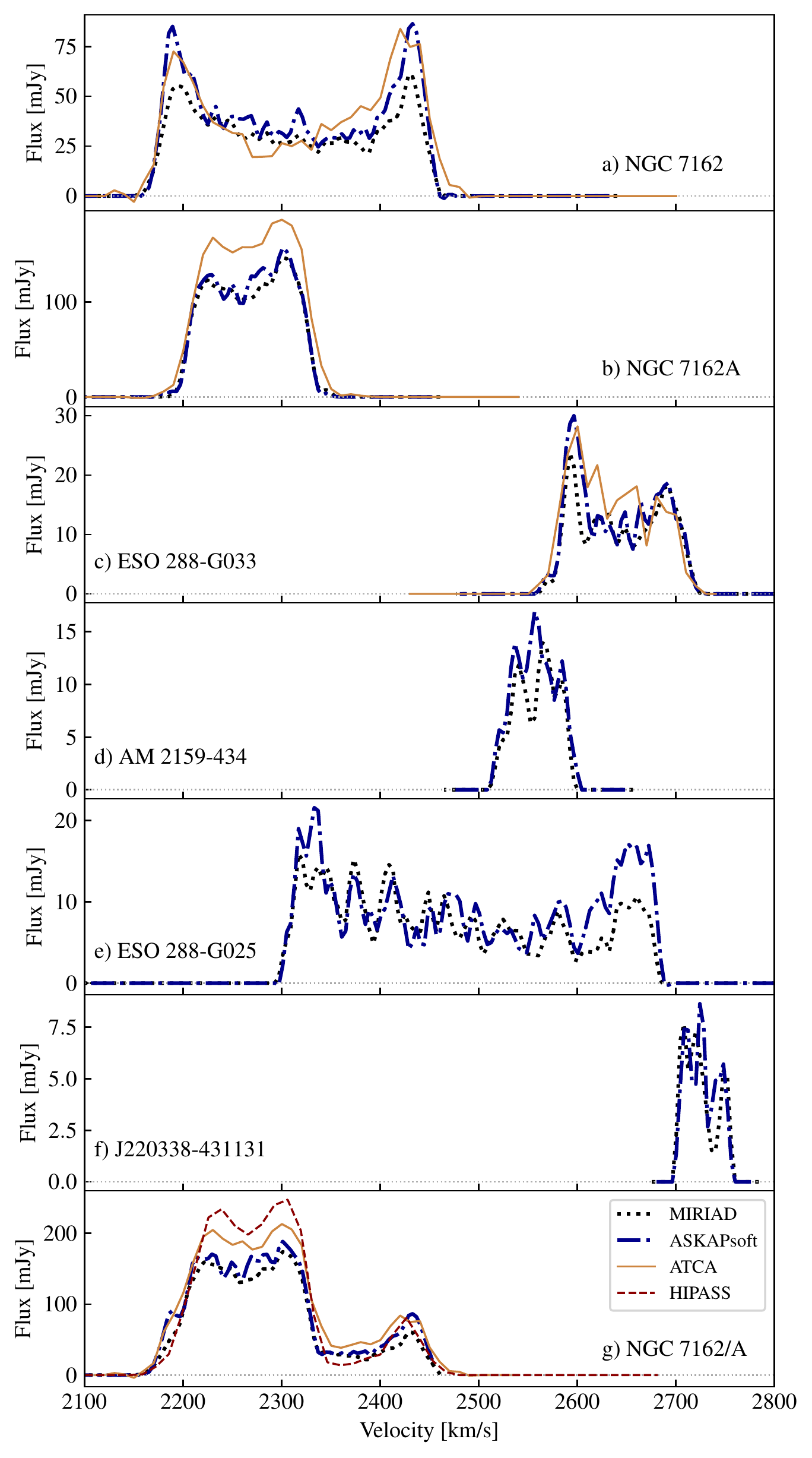}
		\caption{Panels a--f: ASKAP integrated H\,\textsc{i} spectra of our detections as obtained from SoFiA after imaging in both ASKAP\textsc{soft} (thick dot-dashed blue) or \textsc{miriad} (thick dotted black). For comparison we also show ATCA integrated H\,\textsc{i} spectra (thin solid orange). Panel g: The combined ASKAP integrated H\,\textsc{i} spectrum of NGC\,7162 and NGC\,7162A (thick dot-dashed blue) with ATCA (thin solid orange) and HIPASS (thin dashed red) for comparison. We summed the ASKAP/ATCA spectra for NGC\,7162 and NGC\,7162A because these galaxies are not fully separated in the Parkes beam.}
	\label{fig:all_spec}
\end{figure}

One of the main objectives of early science is to compare the ASKAP data against earlier H\,\textsc{i} benchmarks such as ATCA and HIPASS observations. We compare the integrated spectra from SoFiA after processing the ASKAP observations with ASKAP\textsc{soft} and \textsc{miriad} with ATCA (NGC\,7162, NGC\,7162A and ESO\,288-G033) and HIPASS (NGC\,7162 and NGC\,7162A) observations (Fig.~\ref{fig:all_spec}) for validation of the instrument and processing pipeline. We extract galaxy integrated spectra from the \textsc{miriad} imaged ASKAP cube using the SoFiA source mask from the ASKAP\textsc{soft} imaged cube to ensure the spectra cover the same regions. As mentioned in Section~\ref{ss-sec:askap_miriad}, the \textsc{miriad} processing will introduce artefacts (e.g. distortions in source shapes and flux loss) and position errors away from the ASKAP beam centres due to \textsc{miriad} not accounting for non-coplanar baselines. The artefacts and position errors explain the small variations we find between the spectra from the ASKAP\textsc{soft} and \textsc{miriad} imaged data cubes (Fig.~\ref{fig:all_spec}). The Gaussian approximation of the ASKAP beams that we use also contributes to the uncertainty in our measured fluxes as the shape of the edge beams are known to deviate from a two dimensional Gaussian \citep[e.g.][]{Serra2015b,Heywood2016}. The combination of these uncertainties is most notable for the receding side of ESO\,288-G025, closest to the beam edge, in which the spectrum from the \textsc{miriad} cube has only recovered $\sim66\%$ of the flux compared with the ASKAP\textsc{soft} cube between $2600-2700$\,km\,s$^{-1}$ (Fig.~\ref{fig:all_spec}e).

We have good agreement between the integrated ASKAP (ASKAP\textsc{soft}, dot-dashed blue) spectra of NGC\,7162 and ESO\,288-G033 and those we obtain from ATCA (solid orange) observations, where the small flux loss ($\sim6-10\%$) is expected due to the ATCA observations having shorter baselines than ASKAP. We have lost approximately a quarter of the flux of NGC\,7162A in the ASKAP observations, with the larger flux loss due to incomplete deconvolution (Section~\ref{s-sec:maps}). This has also affected NGC\,7162, but only to a small extent compared to NGC\,7162A. We can only compare the spectrum from the sum of the individual NGC\,7162 and NGC\,7162A spectra with HIPASS as the two galaxies are not fully resolved in the Parkes beam. The loss of flux in the combined spectrum relative to HIPASS (dashed red) is again due to the deconvolution process, high source finding threshold and resolving out the H\,\textsc{i} emission with ASKAP.

We can determine velocities of and distances to the two dwarf galaxies using the H\,\textsc{i} spectra. For AM\,2159-434 we derive a value of $V_{\mathrm{hel}}=2558$\,$\mathrm{km}\,\mathrm{s}^{-1}$ (corresponding to $V_{\mathrm{LG}}=2552$\,$\mathrm{km}\,\mathrm{s}^{-1}$ and a luminosity distance of $D_{\mathrm{lum}}=37.9\pm3.0$\,Mpc). Similarly for J220338-431131 we derive $V_{\mathrm{hel}}=2727$\,$\mathrm{km}\,\mathrm{s}^{-1}$ (corresponding to $V_{\mathrm{LG}}=2722$\,$\mathrm{km}\,\mathrm{s}^{-1}$ and $D_{\mathrm{lum}}=40.5\pm3.0$\,Mpc).

% ----------------------------------------------------------------------------------------------------
\subsection{Validation}
\label{s-sec:validation}

\begin{figure}
	\centering
	\includegraphics[width=\columnwidth]{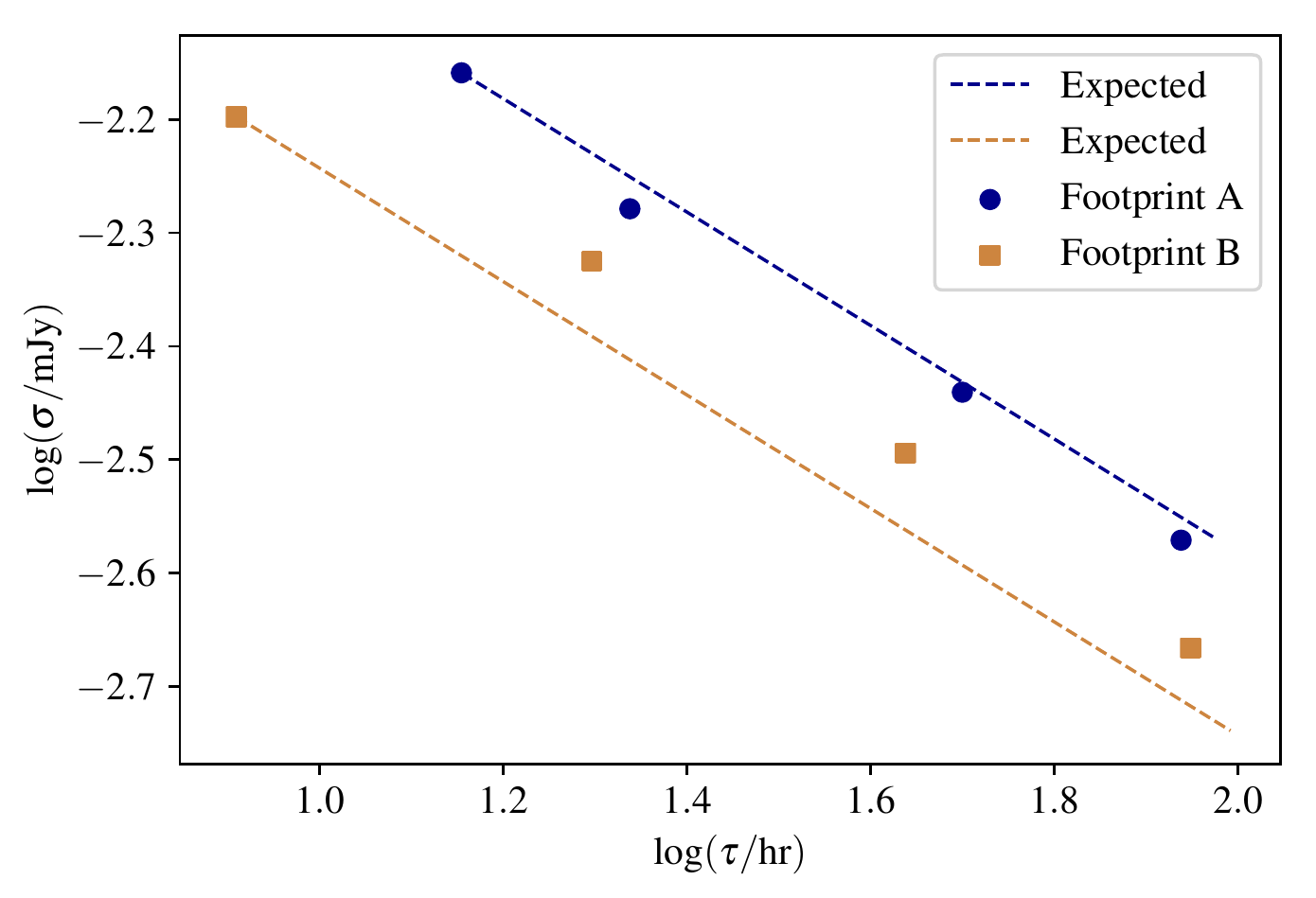}
		\caption{Image RMS noise level as a function of effective integration time $\tau$ for footprint A (blue circle) and footprint B (orange square). The expected RMS (dashed lines) are based on the data. The RMS decreases following the expected relation assuming Gaussian noise ($\sigma\propto1/\sqrt{\tau}$, where $\tau$ is the effective integration time in hours from multiple nights determined as the product the integration time, number of antennas squared and percentage of unflagged data for each night of observation). We only show the effective integration time for 8 nights as this the maximum we have in each footprint and the noise reduction does not scale in the same manner when combining footprints due to primary beam correction.}
	\label{fig:rms_nights}
\end{figure}

We have used the ASKAP observations processed with ASKAP\textsc{soft} for instrument and pipeline verification. We demonstrate that the RMS noise, $\sigma$, in the final ASKAP image data cube decreases as expected assuming Gaussian noise ($\sigma\propto1/\sqrt{\tau}$, where $\tau$ is the effective integration time in hours from multiple nights determined as the product of the integration time, number of antennas squared and percentage of unflagged data for each night of observation). We measure the RMS for footprints A and B separately as the RMS does not scale in the same manner when mosaicking footprints, as the mosaicking process applies primary beam correction (Fig.~\ref{fig:rms_nights}). The RMS in footprint A is measured in the centre of beam 16, one of the edge corner beams, and in footprint B the RMS is measured in the centre of beam 4, a corner beam of the square of beams between the central four beams and edge beams. The sensitivity of the ASKAP PAFs is known to vary as a function of beam position, with the highest sensitivity in the central four beams and the noise increasing towards the edge beams, which we see in the lower RMS in beam 4 vs 16. 

% We do not propagate the integrated flux statistical uncertainties into the calculated H\,\textsc{i} mass uncertainties, which are dominated by the luminosity distance uncertainties

% We measure the integrated flux uncertainties using the statistical uncertainties from the SoFiA signal-to-noise ratio and do not include the larger systematic uncertainties due to data processing and imaging. The calculated H\,\textsc{i} mass uncertainties are dominated by the luminosity distance uncertainties. 

For NGC\,7162, NGC\,7162A and ESO\,288-G033, we compare the parameter output from SoFiA for the ASKAP and ATCA observations (see Table~\ref{table:galaxy_params}). The integrated fluxes and calculated H\,\textsc{i} mass for NGC\,7162 and NGC\,7162A from ASKAP are lower than from ATCA, which is primarily due to flux remaining in residual sidelobes and lack of short baselines. ESO\,288-G033 has reasonable agreement between ASKAP and ATCA with the higher ATCA values primarily due to the shorter baselines in the ATCA configuration providing better sensitivity to diffuse emission. We find very good agreement in $w_{20}$ and $w_{50}$ line profile widths between both observations. We also attempt to compare the ASKAP and ATCA integrated fluxes for NGC\,7162 and NGC\,7162A with their HIPASS values by decomposing the HIPASS moment 0 map into two point sources using the \textsc{miriad} task \textsc{imfit}. The uncertainties in the HIPASS integrated fluxes are only the fitting uncertainties and do not include systematic uncertainties from HIPASS, hence the true uncertainties will be larger. The HIPASS integrated flux for NGC\,7162 is lower, though within errors of the ASKAP and ATCA values, indicating some of its flux has been attributed to NGC\,7162A as we would expect HIPASS to recover more flux (see Table~\ref{table:galaxy_params}). HIPASS recovers more flux in NGC\,7162A than either ASKAP or ATCA indicating that there is likely to be $\sim40\%$ more diffuse H\,\textsc{i} emission in NGC\,7162A, which is resolved out by the interferometers.

\begin{table*}
	\centering
    \caption{NGC\,7162 galaxy group member parameters. $V_{\mathrm{hel}}=cz$ - velocity in the heliocentric reference frame, $V_{\mathrm{LG}}$- velocity in the Local Group reference frame, respectively, $i$ - inclination, $B_{25}$ - 25\,mag\,arcsec$^{-2}$ $B-$band, PA$_{B25}$, $D_{25}$ - position angle and diameter in 25\,mag\,arcsec$^{-2}$ $B-$band, respectively, $D_{\mathrm{lum}}$ - luminosity distance, $\log(L_{X})$ - ROSAT X-ray luminosity, Asym - asymmetry parameter from \textit{Spitzer} IRAC images, $\mathrm{M}_{*}$,  - stellar mass from VISTA Hemisphere Survey, SFR - average star formation rate derived from WISE 12 and 22$\mu$m bands, sSFR - specific star formation rate, $\mathrm{M}_{\mathrm{Dyn}}$ - dynamical mass, $\mathrm{M}_{\mathrm{HI,exp}}$ - expected H\,\textsc{i} mass for morphological type, $S_{\mathrm{int}}$ - integrated flux, $\mathrm{M}_{\mathrm{HI,obs}}$ - observed H\,\textsc{i} mass, $w_{20}$ and $w_{50}$ - observed integrated spectra velocity widths at $20\%$ and $50\%$ of the maximum intensity, PA$_{\mathrm{kin}}$ - kinematic position angle, $\mathrm{DEF}_{\mathrm{HI}}$ - H\,\textsc{i} deficiency, $M_{\mathrm{HI}}/M_*$ - H\,\textsc{i} to stellar mass ratio. References: (1) \protect\cite{Springob2014}, (2) \protect\cite{devaucouleurs1991}, (3) \protect\cite{Lauberts1989}, (4) \protect\cite{Beuing1999}, (5) \protect\cite{Holwerda2014}, (6) \protect\cite{Cluver2017}, (7) This work. Note: The stellar masses for NGC\,7162 and NGC\,7162A are calculated within $<90\arcsec$ and $<105\arcsec$, respectively, not the total stellar mass. The stellar masses for NGC\,7166, ESO\,288-G025 and ESO\,288-G033 are total stellar mass.}
	\label{table:galaxy_params}
	\begin{tabular}{lcccccccr}
		\hline
        \\
		 & NGC\,7162 & NGC\,7162A & NGC\,7166 & ESO\,288-G025 & ESO\,288-G033 & AM\,2159-434 & J220338-431131 & ref \\ 
         \\ \hline
		$\alpha$(J2000) & 21:59:39.1 & 22:00:35.7 & 22:00:32.9 & 21:59:17.9 & 22:02:06.64 & 22:02:50.06 & 22:03:38.62 & (1) \\
		$\delta$(J2000) & -43:18:21 & -43:08:30 & -43:23:23 & -43:52:01 & -43:16:07.0 & -43:26:44.2 & -43:11:28.4 & (1) \\
		Type & SA(s)c & SAB(s)m & SA0 & Scd & Sc & dwarf & dwarf & (2) \\
		$V_{\mathrm{hel}}$ [$\mathrm{km}\,\mathrm{s}^{-1}$] & 2314 & 2271 & 2466 & 2481 & 2641 & 2558 & 2727 & (7) \\
        $V_{\mathrm{LG}}$ [$\mathrm{km}\,\mathrm{s}^{-1}$] & 2309 & 2267 & 2460 & 2473 & 2636 & 2552 & 2722 & (7) \\
		$i$ [deg] & 67.7 & 39.6 & 62.0 & 81.4 & 72.5 & --- & --- & (3) \\
		PA$_{B25}$ [deg] & 10 & 70 & 15 & 53 & 152 & --- & --- & (3) \\
		$B_{25}$ [mag] & 13.44 & 13.79 & 12.87 & 14.01 & 16.66 & --- & --- & (3) \\
		$D_{25}$ [kpc] & 27.65 & 23.61 & --- & 27.39 & 9.83 & --- & --- & (3) \\
        $D_{\mathrm{lum}}$ [Mpc] & $34.3\pm3.0$ & $33.7\pm3.0$ & $33.3\pm3.0$ & $36.7\pm3.0$ & $39.2\pm3.0$ & $37.9\pm3.0$ & $40.5\pm3.0$ & (7) \\
		$\log(L_{X}/\mathrm{erg}\,\mathrm{s}^{-1})$ & --- & --- & $<40.65$ & --- & --- & --- & --- & (4) \\
		Asym (3.6\,$\mu$m) & 0.74 & 0.42 & --- & --- & --- & --- & --- & (5) \\
		Asym (4.5\,$\mu$m) & 0.51 & 0.54 & --- & --- & --- & --- & --- & (5) \\
        $\log(M_{*}/\mathrm{M}_{\sun})$ & $9.7\pm0.1$ & $9.3\pm0.1$ & $10.3\pm0.1$ & $10.0\pm0.1$ & $8.3\pm0.1$ & --- & --- & (7) \\
        SFR [$\mathrm{M}_{\sun}\,\mathrm{yr}^{-1}$] & 0.727 & 0.343 & --- & --- & --- & --- & --- & (6) \\
        $\log(\mathrm{sSFR}/\mathrm{yr})$ & $-9.8$  & $-9.8$ & --- & --- & --- & --- & --- & (7) \\
		$\log(M_{\mathrm{Dyn}}/\mathrm{M}_{\sun})$ & $11.2\pm0.1$ & $10.6\pm0.1$ & --- & $10.9\pm0.1$ & $9.7\pm0.2$ & --- & --- & (7) \\
        $\log(M_{\mathrm{HI,exp}}/\mathrm{M}_{\sun})$ & $8.8\pm0.1$ & $8.8\pm0.1$ & --- & $8.7\pm0.1$ & $8.2\pm0.1$ & --- & --- & (7) \\
        \\
        \textbf{ASKAP} & & & & & & & & \\
        \\
        $S_{\mathrm{int}}$ [Jy\,km\,s$^{-1}$] & $12.37\pm1.24$ & $15.52\pm1.55$ & --- & $3.66\pm0.37$ & $2.04\pm0.20$ & $0.86\pm0.09$ & $0.27\pm0.03$ & (7) \\
        $\log(M_{\mathrm{HI,obs}}/\mathrm{M}_{\sun})$ & $9.5\pm0.1$ & $9.6\pm0.1$ & --- & $9.1\pm0.1$ & $8.9\pm0.1$ & $8.5\pm0.1$ & $8.0\pm0.1$ & (7) \\
		$w_{20}$ [$\mathrm{km}\,\mathrm{s}^{-1}$] & $283\pm4$ & $131\pm4$ & --- & $380\pm4$ & $130\pm4$ & $81\pm4$ & $48\pm4$ & (7) \\
		$w_{50}$ [$\mathrm{km}\,\mathrm{s}^{-1}$] & $269\pm4$ & $119\pm4$ & --- & $362\pm4$ & $109\pm4$ & $60\pm4$ & $45\pm4$ & (7) \\
		PA$_{\mathrm{kin}}$ [deg] & 14 & 28 & --- & 52 & 162 & 15 & 140 & (7) \\
        $\mathrm{DEF}_{\mathrm{HI}}$ & $-0.70$ & $-0.82$ & --- & $-0.39$ & $-0.66$ & --- & --- & (7) \\
        $\log(M_{\mathrm{HI}}/M_*)$ & $-0.2$ & $0.3$ & --- & $-0.9$ & $0.6$ & --- & --- & (7) \\
        \\
        \textbf{ATCA} & & & & & & & \\
        \\
        $S_{\mathrm{int}}$ [Jy\,km\,s$^{-1}$] & $13.20\pm1.32$ & $21.02\pm2.10$ & --- & --- & $2.28\pm0.23$ & --- & --- & (7) \\
        $\log(M_{\mathrm{HI,obs}}/\mathrm{M}_{\sun})$ & $9.6\pm0.1$ & $9.7\pm0.1$ & --- & --- & $8.9\pm0.1$ & --- & --- & (7) \\
		$w_{20}$ [$\mathrm{km}\,\mathrm{s}^{-1}$] & $291\pm10$ & $142\pm10$ & --- & --- & $136\pm10$ & --- & --- & (7) \\
		$w_{50}$ [$\mathrm{km}\,\mathrm{s}^{-1}$] & $272\pm10$ & $119\pm10$ & --- & --- & $106\pm10$ & --- & --- & (7) \\
        $\mathrm{DEF}_{\mathrm{HI}}$ & $-0.72$ & $-0.95$ & --- & --- & $-0.70$ & --- & --- & (7) \\
        \\
        \textbf{HIPASS} & & & & & & & \\
        \\
        $S_{\mathrm{int}}$ [Jy\,km\,s$^{-1}$] & $11.46\pm2.58$ & $23.06\pm2.55$ & --- & --- & --- & --- & --- & (7) \\
		\hline
	\end{tabular}
\end{table*}

% ----------------------------------------------------------------------------------------------------
\section{Rotation Curves}
\label{sec:rotcur}

The advantage of ASKAP observations is the increased resolution (spatial and spectral) and sky coverage compared with archival ATCA and HIPASS data ($39\arcsec\times34\arcsec$, $60\arcsec\times36\arcsec$ and $15.5\arcmin\times15.5\arcmin$, respectively). We can use this to determine the rotation curves for the NGC\,7162 group galaxies. Using rotation curves, we can determine the galaxies' dynamical masses and dark matter content through mass modelling. Due to the different inclinations of the group galaxies (i.e. nearly face-on vs edge-on), we use two methods for determining the rotation curves: $i$) tilted ring fitting to the velocity field \citep[e.g.][Section~\ref{s-sec:tilted_ring}]{Rogstad1974} and $ii$) envelope tracing of the position-velocity diagram along the galaxy's major axis \citep[e.g.][Section~\ref{s-sec:envelope}]{Sancisi1979,Westmeier2013}. The tilted ring modelling method gives more accurate results by allowing for variations in inclination and position angle of the galaxy, but cannot be used for galaxies that are edge-on or are not sufficiently resolved (i.e. ESO\,288-G025, ESO\,288-G033, AM\,2159-434 and J220338-431131). Hence we only use this method for NGC\,7162 and NGC\,7162A. We use the envelope tracing method on all galaxies and in the case of NGC\,7162 and NGC\,7162A we compare the results from the two techniques.

% ----------------------------------------------------------------------------------------------------
\subsection{Tilted Ring Model}
\label{s-sec:tilted_ring}

The tilted ring model works by fitting a series of circular isovelocity rings to the velocity field and assuming that gas particles move with a constant speed on circular orbits. We follow the method of \cite{Westmeier2011} where we use the \textsc{gipsy} \citep{vanderHulst1992,Vogelaar2001} task \textsc{rotcur} and allow the inclination, $i$, position angle, $\theta$, systemic velocity, $v_{\mathrm{sys}}$, and centre position, $(x,y)$, of the rings to vary to best match the galaxy. We use 10 rings, with $\mathrm{radius}=15$\,arcsec, centred on the galaxy, which is around half the ASKAP synthesised beam (this results in some correlation between adjacent points, but improves sampling of the galaxies). 

\begin{figure*}
	\centering
	\includegraphics[width=15cm]{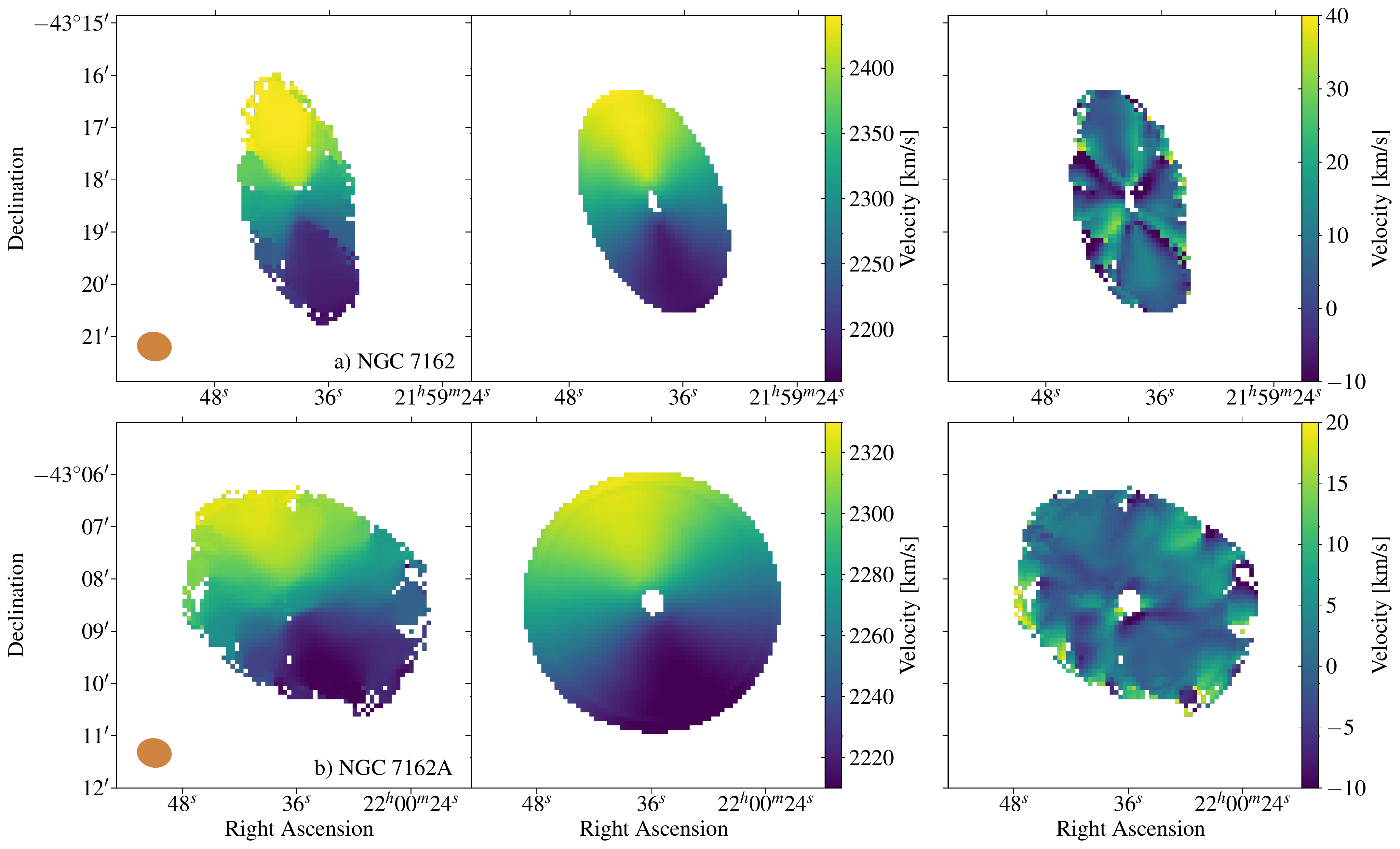}
		\caption{Gauss-Hermite polynomial fit to the observed velocity field, model and residual velocity fields (left, centre and right columns, respectively) for NGC\,7162 and NGC\,7162A (top and bottom rows, respectively). The observed and model velocity fields share the same colour scale. The ASKAP synthesised beam is shown by the orange ellipse in the lower left corner of the left column. The hole in the centre of the model is due to not resolving the central region with the rings with radius of $15\arcsec$.}
	\label{fig:model_fields}
\end{figure*}

\begin{figure*}
	\centering
	\includegraphics[width=15cm]{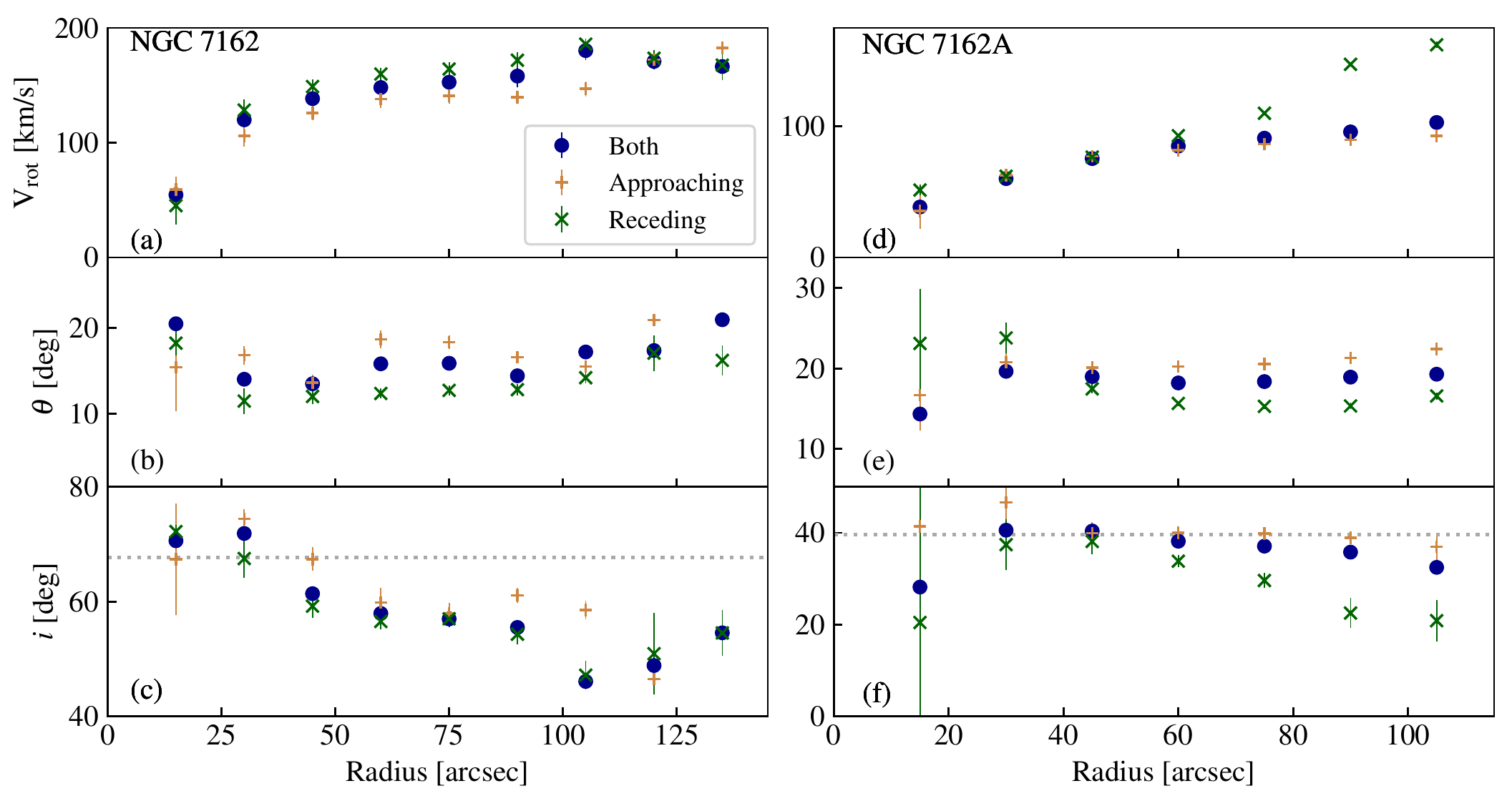}
		\caption{Rotation curve ($\mathrm{V}_{\mathrm{rot}}$), position angle ($\theta$) and inclination angle ($i$) derived from tilted ring modelling of NGC\,7162 and NGC\,7162A (left and right columns, respectively) for both (blue circle), approaching (orange cross) and receding (green `$\times$') sides. The dashed line in panels c and f indicate the inclination angle from $25^{th}$ magnitude $B-$band optical imaging.}
	\label{fig:both_galaxies_kin15}
\end{figure*}

We fit a tilted ring model to both sides of the galaxy together in three iterations. We first leave all parameters free (except the expansion velocity, $v_{\mathrm{exp}}=0\,\mathrm{km}\,\mathrm{s}^{-1}$) and take our initial guess for the galaxy centre from NED. In the second iteration we leave the position angle, inclination angle and rotational velocity free. For the third iteration, we obtain the final rotation curve by leaving only the rotational velocity free. The error in the fitted velocity field is calculated by \textsc{rotcur} using the standard deviation around the mean rotation in each ring. In Fig.~\ref{fig:model_fields}, we show the Gauss-Hermite polynomial fit to the observed velocity field (left column), the model (centre column) and the residual (right column) velocity field for NGC\,7162 and NGC\,7162A (top and bottom rows, respectively). The tilted ring fit provides a good model to the data with small residuals mostly in the range $-10$ to 10\,km\,s$^{-1}$. We also fit the tilted ring model to the approaching and receding sides of the galaxy separately to look for variations in the rotation curve and to estimate the errors in the rotational velocity, position angle and inclination angle (left and right columns of Fig.~\ref{fig:both_galaxies_kin15} for NGC\,7162 and NGC\,7162A, respectively). When fitting each side of each galaxy separately we keep $v_{\mathrm{sys}}$ and $(x,y)$ fixed to the values derived from the fit to the full galaxy.

Both NGC\,7162 and NGC\,7162A have approximately constant position angles across the entire galaxy disk, so neither galaxy has a distinguishable inner and outer disk. The presence of an inner and outer disk could indicate a past interaction event \citep[e.g.][]{Westmeier2011,Westmeier2013}. The inclination of NGC\,7162 decreases towards the edge of the galaxy, while for NGC\,7162A the inclination remains constant. Our tilted ring fits are unaffected by residual sidelobes, as they are below the 5\,$\sigma$ threshold and are beyond the main disks of NGC\,7162 and NGC\,7162A.

There is a degeneracy between the inclination angle, $i$, and rotational velocity, $\mathrm{V}_{\mathrm{rot}}$, best seen for the receding side of NGC\,7162A at large radii ($>75\arcsec$) in Fig.~\ref{fig:both_galaxies_kin15}d and f, where $\mathrm{V}_{\mathrm{rot}}$ rises sharply corresponding to a large decrease in $i$. This degeneracy can explain the differences at large radii between the rotation curves derived from the tilted ring and envelope tracing methods (Fig.~\ref{fig:envelopes}). For NGC\,7162, the higher $\mathrm{V}_{\mathrm{rot}}$ derived from the tilted ring fit is due to the lower $i$ compared with the constant value used for envelope tracing. We left $i$ free in the tilted ring modelling to look for the presence of warps, which we do not find.

We determine the dynamical masses for NGC\,7162 and NGC\,7162A using the rotational velocity determined from the tilted ring fit to the entire galaxy at the last radius before the fitted velocity begins rising sharply between rings (i.e. 130\,$\mathrm{km}\,\mathrm{s}^{-1}$ to 180\,$\mathrm{km}\,\mathrm{s}^{-1}$ at a radius of 135\,arcsec for NGC\,7162A). We find dynamical masses of $\log(M_{\mathrm{dyn}}/\mathrm{M}_{\sun})=11.2\pm0.1$ ($<22.5$\,kpc) and $\log(M_{\mathrm{dyn}}/\mathrm{M}_{\sun})=10.6\pm0.1$ ($<17.2$\,kpc) for NGC\,7162 and NGC\,7162A, respectively.

% ----------------------------------------------------------------------------------------------------
\subsection{Envelope Tracing}
\label{s-sec:envelope}

The envelope tracing method derives a galaxy's rotational velocity by finding the terminal velocity, $v_{\mathrm{t}}$, on the edge of the galaxy facing away from the galaxy's systemic velocity at each position taken along the galaxy's major axis. We follow the method of \cite{Sofue2001,Westmeier2013} for envelope tracing and use position-velocity (PV) diagrams along the kinematic major axis from the SoFiA output (see Fig.~\ref{fig:all_maps4} and \ref{fig:all_maps2}, right hand column, rotational velocities derived from envelope tracing are overlaid in red). We assume that the gas is optically thin when determining the terminal velocity of the gas. This may not be the case in the inner regions and we exclude positions within $\pm20$\arcsec of the galaxy centre in our analysis. We calculate the rotational velocities using equations~1 and 2 from \cite{Westmeier2013}. For consistency, we use a fixed inclination angle for each galaxy from the optical disk (Table~\ref{table:galaxy_params}) for deprojecting the derived rotational velocities.

\begin{figure}
	\centering
	\includegraphics[width=\columnwidth]{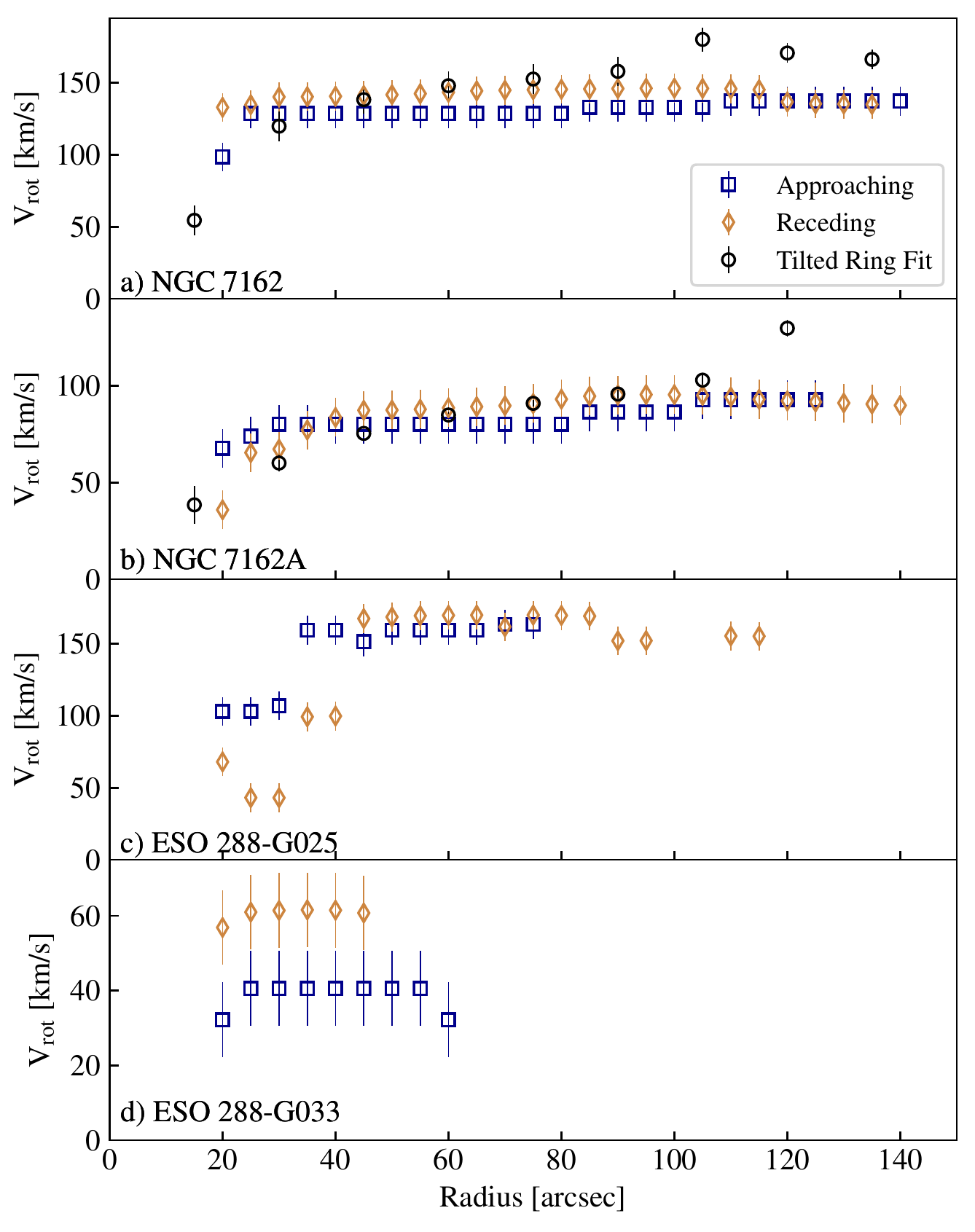}
		\caption{Rotation curves for the approaching (blue squares) and receding (orange diamonds) sides determined from envelope tracing of PV diagrams along the galaxies' kinematic major axis from SoFiA. For NGC\,7162 and NGC\,7162A we plot the rotation curve from the tilted ring model fitting to both sides of the galaxy (black circles), showing good agreement.}
	\label{fig:envelopes}
\end{figure}

We derived rotation curves for the four group spirals using the envelope tracing method. We are unable to derive a rotation curve for AM\,2159-434 or J220338-431131 as we do not have inclination angles, $i$, for either galaxy. Additionally, the PV diagrams for AM\,2159-434 and J220338-431131 are dominated by noise (Fig.~\ref{fig:all_maps2}, right column, rows e and f, respectively). We show our derived rotation curves for the approaching (blue squares) and receding (orange diamonds) sides of each galaxy in Fig.~\ref{fig:envelopes} and show the velocities plotted in red over the PV diagrams in the right hand column of Fig.~\ref{fig:all_maps4}. For NGC\,7162 and NGC\,7162A we also plot the rotation curve from the tilted ring model fitting to both sides of the galaxy (black circles). We find good agreement between the two methods for NGC\,7162 and NGC\,7162A. For NGC\,7162A, we are able to recover the increase in rotational velocity at radii $\leq40\arcsec$ found with the tilted ring analysis using the envelope tracing method, while for NGC\,7162 the envelope tracing method is only able to recover the maximum rotational velocity. The approaching and receding sides of ESO\,288-G025 are in good agreement at radii $>40$\arcsec. ESO\,288-G033, the smallest galaxy in angular size for which we could determine the rotational velocity, also has good agreement at all traced radii.

We determine the dynamical masses for ESO\,288-G025 and ESO\,288-G033 by taking the average of the velocity of the approaching and receding sides of each galaxy at the largest radius at which both are measured (i.e. 80\arcsec and 45\arcsec, respectively). We find dynamical masses of $\log(M_{\mathrm{dyn}}/\mathrm{M}_{\sun})=10.9\pm0.1$ ($<14.2$\,kpc) and $\log(M_{\mathrm{dyn}}/\mathrm{M}_{\sun})=9.7\pm0.2$ ($<9.5$\,kpc) for ESO\,288-G025 and ESO\,288-G033, respectively.

% ----------------------------------------------------------------------------------------------------
\section{Mass Modelling}
\label{sec:mass_mod}

Using the tilted ring derived rotation curves, we model the contributions of the gaseous, stellar and dark matter components to the observed rotation curves of NGC\,7162 and NGC\,7162A, given by
\begin{equation}
	\begin{array}{l}
		\displaystyle v_{\mathrm{rot}}^2(r)=f_{\mathrm{gas}}v_{\mathrm{gas}}^2(r)+f_{\mathrm{*}}v_{\mathrm{*}}^2(r)+v_{\mathrm{dm}}^2(r),
	\end{array}
	\label{equ:rot_cur}
\end{equation}
where $f_{\mathrm{gas}}$ and $f_{\mathrm{*}}$ are mass scaling factors for the gaseous and stellar disks, respectively. We use the \textsc{gipsy} task \textsc{rotmas} to fit to the velocity curve for the three components given in equation~(\ref{equ:rot_cur}), with the gaseous and stellar disk velocities determined from their respective mass surface densities. The mass surface density, $\Sigma(r)$, for each component is used to determine its velocity contribution to the rotational velocity. Our observations have insufficient resolution of the inner disk region to differentiate among various dark matter density profiles (e.g. pseudo-isothermal, NFW). However, we do provide estimates of the dark matter mass required for the observed rotation curve.

% --------------------------------------------------------------------
\subsection{Gas Component}
\label{s-sec:gas_comp}

The gas surface density is derived using the H\,\textsc{i} column density profile, $N_{\mathrm{HI}}(r)$, corrected for inclination, $i$, taken from the tilted ring fit with
\begin{equation}
	\begin{array}{l}
		\displaystyle \Sigma_{\mathrm{gas}}(r)=fm_{\mathrm{H}}N_{\mathrm{HI}}(r)\cos(i),
	\end{array}
\end{equation}
where the hydrogen atom mass is $m_{\mathrm{H}}=1.674\times10^{-27}$\,kg and $f$ is a correction factor accounting for the contribution of helium in the disk \citep[we assume $f=1.4$ for the contribution of helium, e.g.][]{Westmeier2011}. We use the \textsc{gipsy} task \textsc{ellint} to calculate the H\,\textsc{i} surface density profile, which we scale by $f$, from the H\,\textsc{i} column density map, the total H\,\textsc{i} gas mass and the luminosity distance of the galaxy. We assume the H\,\textsc{i} gas is optically thin and is in an infinitely thin disk. The gas surface density will be underestimated if dense locations in the disk are not optically thin. The mass contribution of molecular and ionised gas can be accounted for using the scaling factor $f_{\mathrm{gas}}$ in equation~\ref{equ:rot_cur}.

We also scaled the total gas mass of NGC\,7162A by a factor of 1.25 (i.e. to the mass from ATCA) to approximately account for the flux/mass loss due to incomplete deconvolution of the ASKAP data ($\sim5.5$\,Jy\,km\,s$^{-1}$, Fig.~\ref{fig:all_spec}b). We scaled to the ATCA flux rather than HIPASS due to the confusion in the HIPASS detection of NGC\,7162 and NGC\,7162A. However, ATCA appears to have recovered the majority of the flux from comparison of the combined spectra of NGC\,7162 and NGC\,7162A from ATCA with the HIPASS spectrum (Fig.~\ref{fig:all_spec}g). We did not apply this correction to NGC\,7162 as it only suffers a small flux loss ($\sim1$\,Jy\,km\,s$^{-1}$, Fig.~\ref{fig:all_spec}a). The incomplete deconvolution most strongly affects the recovered flux of NGC\,7162A as it is the brightest source with more flux left in sidelobes detectable above the noise. In Fig.~\ref{fig:surface_density}, we show the H\,\textsc{i} mass surface density with blue squares for NGC\,7162 and NGC\,7162A.

% --------------------------------------------------------------------
\subsection{Stellar Component}
\label{s-sec:star_comp}

The stellar surface density is determined by converting the optical or near-infrared flux density, $S_{\lambda}(r)$, to stellar mass surface density, $\Sigma_{\mathrm{stellar}}(r)$, using the stellar mass-to-light ratio, $\Upsilon_{\lambda}$. Both the flux density and mass-to-light ratio are wavelength dependent and are determined for a specific photometric band \citep[e.g. IRAC 3.6 and 4.5\,$\mu$m bands,][]{Westmeier2011}. The mass-to-light ratio conversion uses solar units and takes the form
\begin{equation}
	\begin{array}{l}
		\displaystyle \Sigma_{\mathrm{stellar}}(r)\sim \Upsilon_{\lambda}S_{\lambda}(r).
	\end{array}
\end{equation}
We use the VISTA Hemisphere Survey \citep[VHS,][]{McMahon2013} $J-$ and $K-$band images for deriving stellar masses and surface densities. We determine stellar masses for VHS $J-$ and $K-$band images using masses estimated from the Galaxy and Mass Assembly Survey \citep[GAMA,][]{Driver2011} and VISTA Kilo-Degree Infrared Galaxy Survey (VIKINGs) absolute magnitudes \citep{Wright2016} for $\sim90000$ galaxies. The stellar masses determined from the GAMA survey are tightly correlated (scatter of 0.1\,dex) with the $J-$ and $K-$band absolute magnitudes, with the relations given by 
\begin{equation}
	\begin{array}{l}
		\displaystyle \log_{10}(M_{*}/\mathrm{M}_{\sun})=-0.454(J_{\mathrm{abs}})+0.384
	\end{array}
    \label{equ:gama_vhs_j}
\end{equation}
and
\begin{equation}
	\begin{array}{l}
		\displaystyle \log_{10}(M_{*}/\mathrm{M}_{\sun})=-0.407(K_{\mathrm{abs}})+1.32,
	\end{array}
    \label{equ:gama_vhs_k}
\end{equation}
where $J_{\mathrm{abs}}$ and $K_{\mathrm{abs}}$ are the absolute magnitude in the $J-$ and $K-$bands, respectively. We can then use equations~\ref{equ:gama_vhs_j} and \ref{equ:gama_vhs_k} and the conversion from VHS image flux units, $A$, to magnitudes ($\mathrm{mag}=30-2.5\log(A)$, 30 is the zero point magnitude) to derive linear relationships between stellar mass and the $J-$ and $K-$band image flux units
 \begin{equation}
	\begin{array}{l}
		\displaystyle M_{*}=0.013D_{\mathrm{lum}}^{2.27}A_{J}^{1.135}\,\mathrm{M}_{\sun}
	\end{array}
    \label{equ:gama_vhs_j_linear}
\end{equation}
and
\begin{equation}
	\begin{array}{l}
		\displaystyle M_{*}=0.207D_{\mathrm{lum}}^{2.04}A_{K}^{1.02}\,\mathrm{M}_{\sun},
	\end{array}
    \label{equ:gama_vhs_k_linear}
\end{equation}
where $D_{\mathrm{lum}}$ is the luminosity distance. We then calculate the stellar mass from the $J-$ and $K-$band images by summing the image flux in rings with the position angle and inclination of each ring determined from the tilted ring modelling for NGC\,7162 and NGC\,7162A (for consistency with our H\,\textsc{i} rings) and using a fixed position angle and inclination, from SoFiA and optical imaging, respectively, for ESO\,288-G025 and ESO\,288-G033. For NGC\,7162 and NGC\,7162A, we calculate the stellar mass contained within a given radius ($<90\arcsec$ and $<105\arcsec$, respectively), corresponding to the maximum radius at which we determine the stellar surface densities (i.e. where the surface density levels off at $<0.1\,\mathrm{M}_{\sun}\,\mathrm{pc}^{-2}$). We again use the \textsc{gipsy} task \textsc{ellint}, this time to derive both the stellar masses for NGC\,7162, NGC\,7162A, ESO\,288-G025 and ESO\,288-G033 and the stellar surface density profiles for NGC\,7162 and NGC\,7162A. We do not calculate a stellar mass for the dwarf galaxies due to foreground stars in the VHS images. Our final quoted stellar masses in Table~\ref{table:galaxy_params} are the average of the masses found in the $J-$ and $K-$bands, similarly for the stellar surface density profiles of NGC\,7162 and NGC\,7162A (Fig.~\ref{fig:surface_density}). Differences in the surface density profiles derived in each band are smaller than the uncertainties. We note there are a number of factors creating uncertainty in the calculated masses including the level of dust extinction, the galaxy's initial mass function \citep[][IMF used for GAMA]{Chabrier2003}, star formation history and metallicity, in addition to radial variation of these parameters within the disk \citep{Oh2008,Westmeier2011}. See \cite{Taylor2011} for details on derivation process and assumptions used in deriving GAMA masses.

We note that total stellar masses of $\log(M/\mathrm{M}_{\sun})=10.194$ and 9.812 for NGC\,7162 and NGC\,7162A, respectively, were calculated from the S$^4$G survey on \textit{Spitzer} \citep{Munoz2015}. The \textit{Spitzer} derived stellar masses are higher than the VHS values due to our radial cutoffs in measured values and the different IMFs assumed, \cite{Salpeter1955} and \cite{Chabrier2003}, respectively. We can convert the stellar masses between IMFs using equation~12 from \cite{Longhetti2009},
\begin{equation}
	\begin{array}{l}
		\displaystyle M_{\mathrm{Chabrier}}=0.55\times M_{\mathrm{Salpeter}}
    \end{array}
    \label{equ:convert_imf}
\end{equation}
and extend the radius to which we measure the stellar mass, bringing the two values into agreement, within errors ($\log[M_{\mathrm{Chabrier,Spitzer}}/\mathrm{M}_{\sun}]=9.9$ and 9.6 and $\log[M_{\mathrm{VHS,total}}/\mathrm{M}_{\sun}]=9.7$ and 9.5, for NGC\,7162 and NGC\,7162A, respectively). We use VHS rather than \textit{Spitzer} data to provide consistency in the derived stellar masses of the group spirals as there are no \textit{Spitzer} observations of ESO\,288-G025 and ESO\,288-G033. Additionally, VHS covers nearly the entire Southern hemisphere and will be complementary to WALLABY for providing stellar maps.

% --------------------------------------------------------------------
\subsection{Vertical Density}
\label{ss-sec:vertical}

\begin{figure}
	\centering
	\includegraphics[width=\columnwidth]{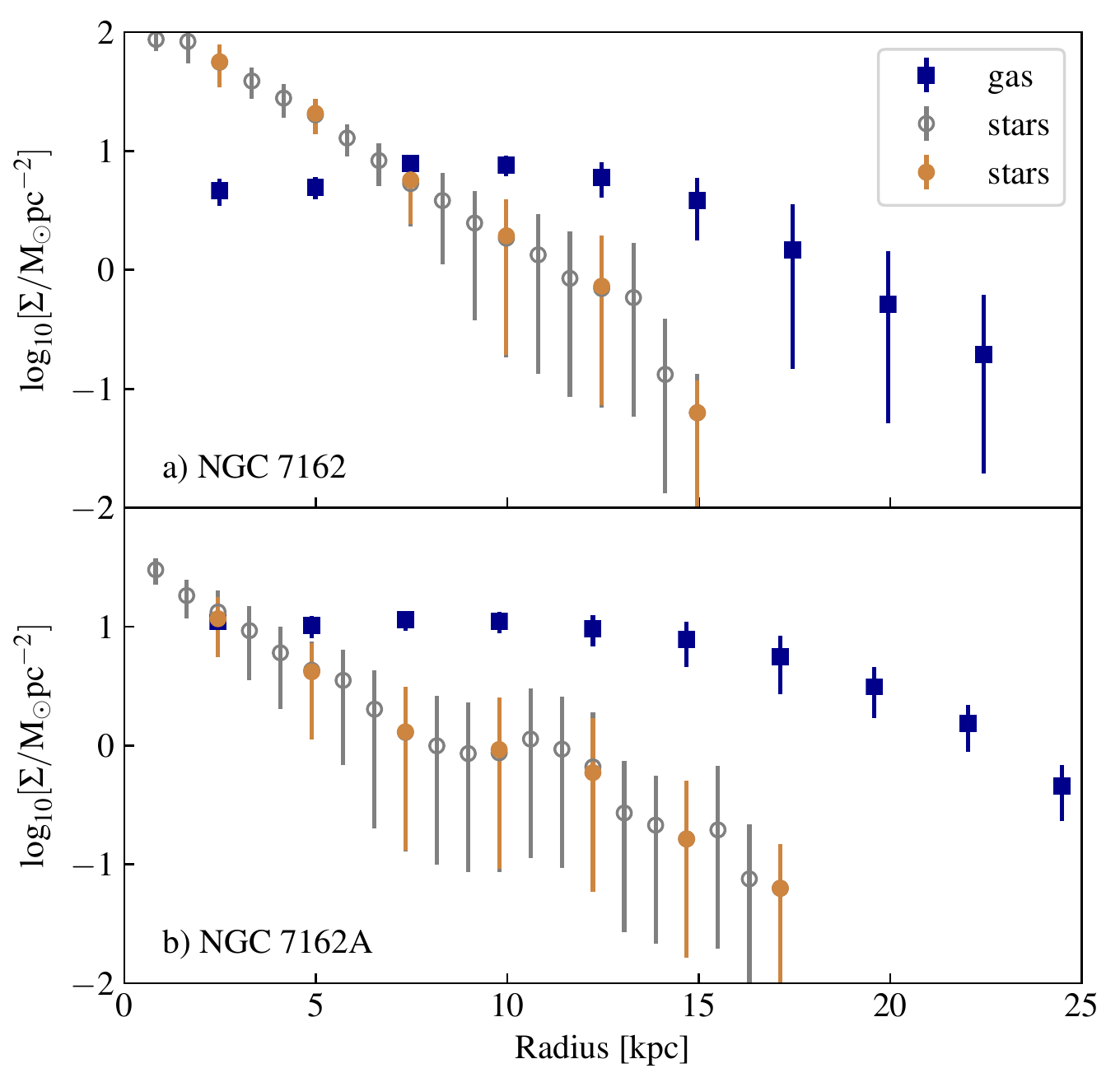}
		\caption{Stellar (filled, orange circles) and gaseous (filled, blue squares) surface densities for NGC\,7162 (panel a) and NGC\,7162A (panel b) using 15\arcsec width rings. The stellar surface density determined using 5\arcsec width rings is shown with empty, grey circles. The stellar surface densities are the average of the $J-$ and $K-$band surface densities.}
	\label{fig:surface_density}
\end{figure}

For both the gaseous and stellar disks, we must consider the vertical density profile, $\rho(z)$, although there is no consistent method used for their modelling \citep[e.g.][and references therein]{Westmeier2011}. The most commonly used vertical density profiles use either infinitely thin gas and/or stellar disks or a density distribution with a $\mathrm{sech}^2(z/z_0)$ vertical dependence \citep[based on studies of edge-on spiral galaxies by][]{Kruit1981b,Kruit1981a}:
\begin{equation}
	\begin{array}{l}
		\displaystyle \rho(r,z)=\rho(r)\mathrm{sech}^2(z/z_0),
	\end{array}
	\label{equ:vertical_density}
\end{equation}
where $z_0$ is the scale height of the disk. In this work, we model the gas as an infinitely thin disk and the stellar disk vertical density following equation~\ref{equ:vertical_density} and assume a scale length-to-scale height ratio of $h/z_0=5$ \citep[e.g.][]{Kruit1981b}. We calculate the scale length, $h$, using stellar surface density profiles determined using $5\arcsec$ rings, which better constrains the radial variation in the stellar surface density with higher resolution in the inner region of the galaxy disks. NGC\,7162 has scale length of $h=6.57\pm0.05$\,kpc and a scale height of $z_0=1.31\pm0.05$\,kpc. Likewise, NGC\,7162A has a scale length of $h=4.65\pm0.05$\,kpc and a scale height of $z_0=0.93\pm0.05$\,kpc. We obtain the same stellar surface density profile (average of $J-$ and $K-$bands) for NGC\,7162 and NGC\,7162A using either $5\arcsec$ and $15\arcsec$ rings (Fig.~\ref{fig:surface_density}, empty grey and filled orange circles, respectively). We also considered the case of the gas disk having the same scale height as the stellar disk. However, this has only a small affect of lowering the gas velocity by $\sim1$\,km\,s$^{-1}$ at all radii and increasing the derived dark matter mass by $\sim0.1$\,dex.

\begin{table*}
	\centering
    \caption{NGC\,7162 tilted ring model fit parameters and their errors, and gaseous and stellar mass surface densities and their RMS deviations. Parameters: $r$ - radius, $i$ - inclination, $\theta$ - position angle, $v_{\mathrm{rot}}$ - rotational velocity, $\Sigma_{*}^{J}$ - $J-$band stellar mass surface density, $\Sigma_{*}^{K}$ - $K-$band stellar mass surface density, $\Sigma_{*}$ - average stellar mass surface density, $\Sigma_{\mathrm{gas}}$ - gas mass surface density.}
	\label{table:ngc7162_fit_params}
	\begin{tabular}{ccccccccc}
		\hline
		 $r$ & $r$ & $i$ & $\theta$ & $v_{\mathrm{rot}}$ & $\Sigma_{*}^{J}$ & $\Sigma_{*}^{K}$ & $\Sigma_{*}$ & $\Sigma_{\mathrm{gas}}$ \\
         {[arcsec]} & [kpc] & [degrees] & [degrees] & [$\mathrm{km}\,\mathrm{s}^{-1}$] & [$\mathrm{M}_{\sun}\,\mathrm{pc}^{-2}$] & [$\mathrm{M}_{\sun}\,\mathrm{pc}^{-2}$] & [$\mathrm{M}_{\sun}\,\mathrm{pc}^{-2}$] & [$\mathrm{M}_{\sun}\,\mathrm{pc}^{-2}$] \\ \hline
		15 & 2.49 & $71\pm4$ & $20\pm2$ & $54\pm11$ & $47.4\pm25.8$ & $65.3\pm36.8$ & $56.4\pm22.1$ & $4.6\pm1.2$ \\ 
		30 & 4.99 & $72\pm2$ & $14\pm1$ & $120\pm10$ & $18.3\pm8.0$ & $23.2\pm11.4$ & $20.8\pm6.9$ & $5.0\pm1.1$ \\ 
		45 & 7.48 & $61\pm2$ & $13\pm1$ & $138\pm10$ & $5.5\pm3.2$ & $5.9\pm6.0$ & $5.7\pm3.3$ & $7.8\pm1.0$ \\ 
		60 & 9.98 & $58\pm2$ & $16\pm1$ & $148\pm10$ & $2.0\pm1.8$ & $1.9\pm3.8$ & $1.9\pm2.0$ & $7.6\pm1.5$ \\ 
		75 & 12.47 & $57\pm2$ & $16\pm1$ & $153\pm10$ & $0.8\pm1.6$ & $0.7\pm1.9$ & $0.7\pm1.2$ & $6.0\pm2.0$ \\ 
		90 & 14.96 & $55\pm2$ & $14\pm1$ & $158\pm10$ & $0.1\pm0.1$ & $0.1\pm0.1$ & $0.1\pm0.1$ & $3.9\pm2.1$ \\ 
		105 & 17.45 & $46\pm5$ & $17\pm1$ & $180\pm8$ & --- & --- & --- & $1.5\pm2.1$ \\ 
		120 & 19.94 & $49\pm3$ & $17\pm1$ & $171\pm7$ & --- & --- & --- & $0.5\pm0.9$ \\ 
		135 & 22.44 & $55\pm4$ & $21\pm2$ & $166\pm7$ & --- & --- & --- & $0.2\pm0.4$ \\ 
		\hline
	\end{tabular}
\end{table*}

\begin{table*}
	\centering
    \caption{NGC\,7162A tilted ring model fit parameters and gaseous and stellar mass surface densities. Parameters the same as Table~\ref{table:ngc7162_fit_params}.}
	\label{table:ngc7162a_fit_params}
	\begin{tabular}{ccccccccc}
		\hline
		 $r$ & $r$ & $i$ & $\theta$ & $v_{\mathrm{rot}}$ & $\Sigma_{*}^{J}$ & $\Sigma_{*}^{K}$ & $\Sigma_{*}$ & $\Sigma_{\mathrm{gas}}$ \\
         {[arcsec]} & [kpc] & [degrees] & [degrees] & [$\mathrm{km}\,\mathrm{s}^{-1}$] & [$\mathrm{M}_{\sun}\,\mathrm{pc}^{-2}$] & [$\mathrm{M}_{\sun}\,\mathrm{pc}^{-2}$] & [$\mathrm{M}_{\sun}\,\mathrm{pc}^{-2}$] & [$\mathrm{M}_{\sun}\,\mathrm{pc}^{-2}$] \\ \hline
		15 & 2.45 & $34\pm11$ & $18\pm6$ & $38\pm10$ & $9.7\pm5.9$ & $13.6\pm11.5$ & $11.7\pm6.2$ & $11.0\pm2.3$ \\ 
		30 & 4.89 & $44\pm5$ & $21\pm2$ & $60\pm4$ & $3.6\pm2.4$ & $4.7\pm6.2$ & $4.2\pm3.1$ & $10.2\pm2.2$ \\ 
		45 & 7.34 & $40\pm2$ & $19\pm1$ & $75\pm2$ & $1.2\pm1.4$ & $1.4\pm3.7$ & $1.3\pm1.8$ & $11.4\pm2.2$ \\ 
		60 & 9.79 & $38\pm1$ & $18\pm1$ & $85\pm2$ & $0.7\pm1.2$ & $1.1\pm3.3$ & $0.9\pm1.6$ & $11.1\pm2.2$ \\ 
		75 & 12.23 & $37\pm1$ & $18\pm1$ & $91\pm2$ & $0.6\pm1.2$ & $0.6\pm2.0$ & $0.6\pm1.1$ & $9.6\pm2.8$ \\ 
		90 & 14.68 & $36\pm2$ & $19\pm1$ & $96\pm2$ & $0.3\pm0.7$ & $0.1\pm0.3$ & $0.2\pm0.3$ & $7.8\pm3.2$ \\ 
		105 & 17.13 & $34\pm3$ & $19\pm1$ & $103\pm3$ & $0.1\pm0.1$ & $0.1\pm0.1$ & $0.1\pm0.1$ & $5.6\pm2.9$ \\ 
		120 & 19.59 & $26\pm6$ & $19\pm1$ & $130\pm4$ & --- & --- & --- & $3.1\pm1.4$ \\ 
		135 & 22.03 & $18\pm26$ & $20\pm4$ & $182\pm6$ & --- & --- & --- & $1.5\pm0.7$ \\ 
		150 & 24.48 & $11\pm148$ & $16\pm16$ & $319\pm7$ & --- & --- & --- & $0.5\pm0.2$ \\ 
		\hline
	\end{tabular}
\end{table*}

% --------------------------------------------------------------------
\subsection{Dark Matter Component}
\label{s-sec:dm_comp}

The third component to the mass model is the dark matter halo, which can be modelled following a number of different profiles. In this work, we only use a pseudo-isothermal profile as we are unable to differentiate among different dark matter profiles due to the low resolution of the central region of NGC\,7162 and NGC\,7162A. The pseudo-isothermal model \citep[e.g.][]{Begeman1991} is a physically motivated model with a constant central density with the density profile given by,
\begin{equation}
	\begin{array}{l}
		\displaystyle \rho(r)=\frac{\rho_0}{1+(r/r_{\mathrm{c}})^2},
	\end{array}
\end{equation}
where $\rho_0$ is the central density and $r_{\mathrm{c}}$ is the core radius. The pseudo-isothermal velocity profile is given by,
\begin{equation}
	\begin{array}{l}
		\displaystyle v^2(r)=4\pi \mathrm{G}\rho_0 r_{\mathrm{c}}^2 \left[1+\frac{r_{\mathrm{c}}}{r}\arctan\left(\frac{r}{r_{\mathrm{c}}}\right)\right].
	\end{array}
\end{equation}

% --------------------------------------------------------------------
\subsection{Modelling}
\label{s-sec:modelling}

We use the \textsc{gispy} task \textsc{rotmas} for our mass modelling of NGC\,7162 and NGC\,7162A. The inputs into \textsc{rotmas} are the observed rotation curve from the tilted ring fit to the entire galaxy, the gaseous and stellar surface densities, our selected vertical stellar density distribution given by equation~\ref{equ:vertical_density}, the stellar disk scale height and our chosen dark matter density profile (pseudo-isothermal). We do not include a bulge component in our modelled stellar surface density profile as, according to their morphological classifications (Table~\ref{table:galaxy_params}), the galaxies do not have prominent bulges. We fix the gaseous and stellar scaling factors, $f_{\mathrm{gas}}$ and $f_{*}$, to unity, leaving only the dark matter profile parameters, $\rho_0$ and $r_{\mathrm{c}}$, free. We also tested leaving $f_{\mathrm{gas}}$ as a free parameter, however this did not significantly alter the derived dark matter masses. For NGC\,7162 and NGC\,7162A, we fit out to the same radius to which we calculated dynamical masses (20.3 and 15.5\,kpc, respectively). We perform the mass modelling for a single dark matter profile with fixed gaseous and stellar scaling factors because of our limited resolution and inability to resolve the galaxies' central region (i.e. where we would be able to distinguish between different models). 

\begin{figure}
	\centering
	\includegraphics[width=\columnwidth]{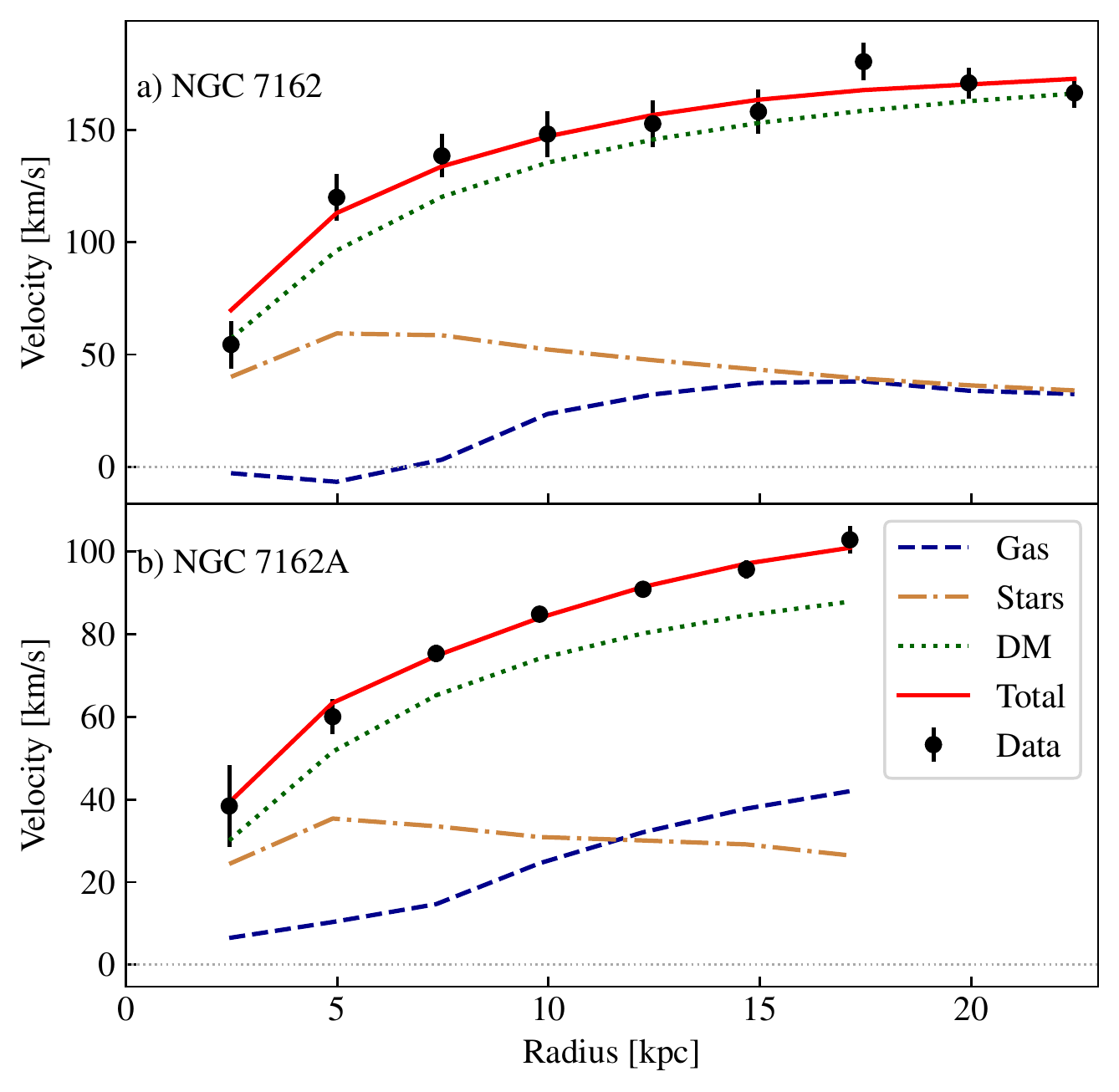}
		\caption{Rotation curve contributions for gas (dashed blue), stars (dot-dashed orange), dark matter (dotted green) and total (solid red) from mass modelling with \textsc{rotmas} for NGC\,7162 (panel a) and NGC\,7162A (panel b). Also plotted is the observed rotation curve (black points).}
	\label{fig:vel_profile}
\end{figure}

In Fig.~\ref{fig:vel_profile}, we show our mass models for NGC\,7162 (panel a) and NGC\,7162A (panel b). We note the negative velocities of the gas at radii $<7.5$\,kpc in NGC\,7162 do not mean the gas is counter-rotating to the rest of the disk, but is due to test particles in the modelling having a net outward force resulting in a negative $v_{\mathrm{gas}}^2$ \citep{Westmeier2011}. This gives an imaginary velocity value, which is represented by negative velocities. Our derived dark matter masses are in agreement with our results from estimating dynamical masses and show both galaxies are dark matter dominated ($>81\,\%$, Table~\ref{table:dm_params}).

\begin{table}
	\centering
    \caption{Mass modelling parameters using a pseudo-isothermal dark matter density profile for NGC\,7162 and NGC\,7162A. $\rho_0$ is the central density, $r_{\mathrm{c}}$ is the core radius, $\chi_{\mathrm{red}}^2$ is the reduced $\chi^2$ goodness of fit, $\log(M_{\mathrm{DM}}/\mathrm{M}_{\sun})$ is the dark matter mass and $f_{\mathrm{DM}}$ is the dark matter fraction.}
	\label{table:dm_params}
	\begin{tabular}{lcccr}
		\hline
		ISO Parameter & NGC\,7162 &  NGC\,7162A \\ \hline
        $r_{\mathrm{c}}$ [kpc] & $4.09\pm0.70$  & $4.35\pm0.39$  \\
		$\rho_{0}$ [$\mathrm{M}_{\sun}\,\mathrm{pc}^{-3}$] &  $41.90\pm10.95$ &  $11.80\pm1.36$ \\
		$\chi_{\mathrm{red}}^2$ & 0.91  & 0.41  \\
		$\log(M_{\mathrm{DM}}/\mathrm{M}_{\sun})$ & $11.1\pm0.2$  & $10.4\pm0.2$  \\
		$f_{\mathrm{DM}}$ &  0.95 & 0.81  \\
		\hline
	\end{tabular}
\end{table}

% ----------------------------------------------------------------------------------------------------
\section{HI Gas Mass and Deficiency}
\label{sec:hi_gas}

Following \cite{Haynes1984} and \cite{Cortese2011}, we determine if the group spirals are H\,\textsc{i} deficient, normal or rich compared to similar field spirals by calculating the expected galaxy H\,\textsc{i} mass as a function of galaxy morphology and size, 
\begin{equation}
	\begin{array}{l}
		\displaystyle \log(M_{\mathrm{HI,exp}}/\mathrm{M}_{\sun})=a_{\mathrm{HI}} + b_{\mathrm{HI}} \times \log\left(\frac{hD_{25}}{\mathrm{kpc}}\right) - 2\log(h),
	\end{array}
\end{equation}
where $h=H_0/100\,\mathrm{km\,s}^{-1}\,\mathrm{Mpc}^{-1}$, $D_{25}$ is the optical 25\,mag\,arcsec$^{-2}$ $B-$band diameter and $a_{\mathrm{HI}}$ and $b_{\mathrm{HI}}$ are morphological type dependent coefficients \citep[see table~3 from][]{Boselli2009}. The H\,\textsc{i} deficiency is then defined to be,
\begin{equation}
	\begin{array}{l}
		\displaystyle \mathrm{DEF}_{\mathrm{H\,\textsc{i}}} = \log(M_{\mathrm{HI,exp}}/\mathrm{M}_{\sun}) - \log(M_{\mathrm{HI,obs}}/\mathrm{M}_{\sun}).
	\end{array}
\end{equation}

We obtain morphology classifications and calculate galaxy diameters from NED (Table~\ref{table:galaxy_params}). We used the H\,\textsc{i} masses from ATCA for NGC\,7162 and NGC\,7162A for calculating their deficiencies as our ATCA spectra agree with HIPASS and recover more of the flux than ASKAP-12, hence we are less likely to underestimate $\mathrm{DEF}_{\mathrm{H\,\textsc{i}}}$. We calculate the H\,\textsc{i} deficiencies for ESO\,288-G025 and ESO\,288-G033 using the H\,\textsc{i} masses from ASKAP, as these galaxies do not have the same issues.

NGC\,7162, NGC\,7162A and ESO\,288-G033 have H\,\textsc{i} excesses of 0.72, 0.95 and 0.66\,dex, respectively. The excesses in these galaxies indicate that they have either accreted additional gas or are yet to be affected by the group environment, which would generally cause galaxies near the group centre to lose gas and become H\,\textsc{i} deficient. ESO\,288-G025 has a much lower excess of 0.33\,dex. \cite{Kilborn2009} and \cite{Rasmussen2012} only consider galaxies to have a H\,\textsc{i} deficiency if |DEF$_{\mathrm{HI}}$| $>0.30$ and $>0.45$, respectively, to account for uncertainties in diameters, H\,\textsc{i} masses and galaxy morphologies causing the apparent deficiencies. Here we take a conservative value of only considering galaxies with |DEF$_{\mathrm{HI}}|>0.45$ having a H\,\textsc{i} deficiency or excess. Hence, we do not consider ESO\,288-G025 to have a H\,\textsc{i} excess.

% -------------------------------------------------------------------------------------------------------------------------------------------------------------------------
\section{DISCUSSION}
\label{sec:discussion}

In this work, we search for signs of interactions within the NGC\,7162 group, primarily between NGC\,7162 and NGC\,7162A. However, we do not identify any morphological signs of interactions or extended H\,\textsc{i} emission around NGC\,7162 or NGC\,7162A at ASKAP-12 resolution and sensitivity. We could expect to find evidence of interactions as the largest and approximately central galaxy of this group, NGC\,7166, is an early type elliptical galaxy, suggesting that one or more merger events likely occurred in the group's history. It is possible that the merger occurred sufficiently long in the past that most signs have disappeared as there is no H\,\textsc{i} detected around NGC\,7166 and all detected gas is confined to the six detected galaxies. We can place an upper limit on the column density of any extra-planar gas of $<9.5\times10^{18}$\,cm$^{-2}$ per 4\,km\,s$^{-1}$ channel for emission filling the $39\arcsec\times34\arcsec$ beam using our 5\,$\sigma\sim11.5$\,mJy\,beam$^{-1}$ threshold. However, we can look for internal indications of interactions within the galaxies in H\,\textsc{i}.

\begin{figure}
	\centering
	\includegraphics[width=\columnwidth]{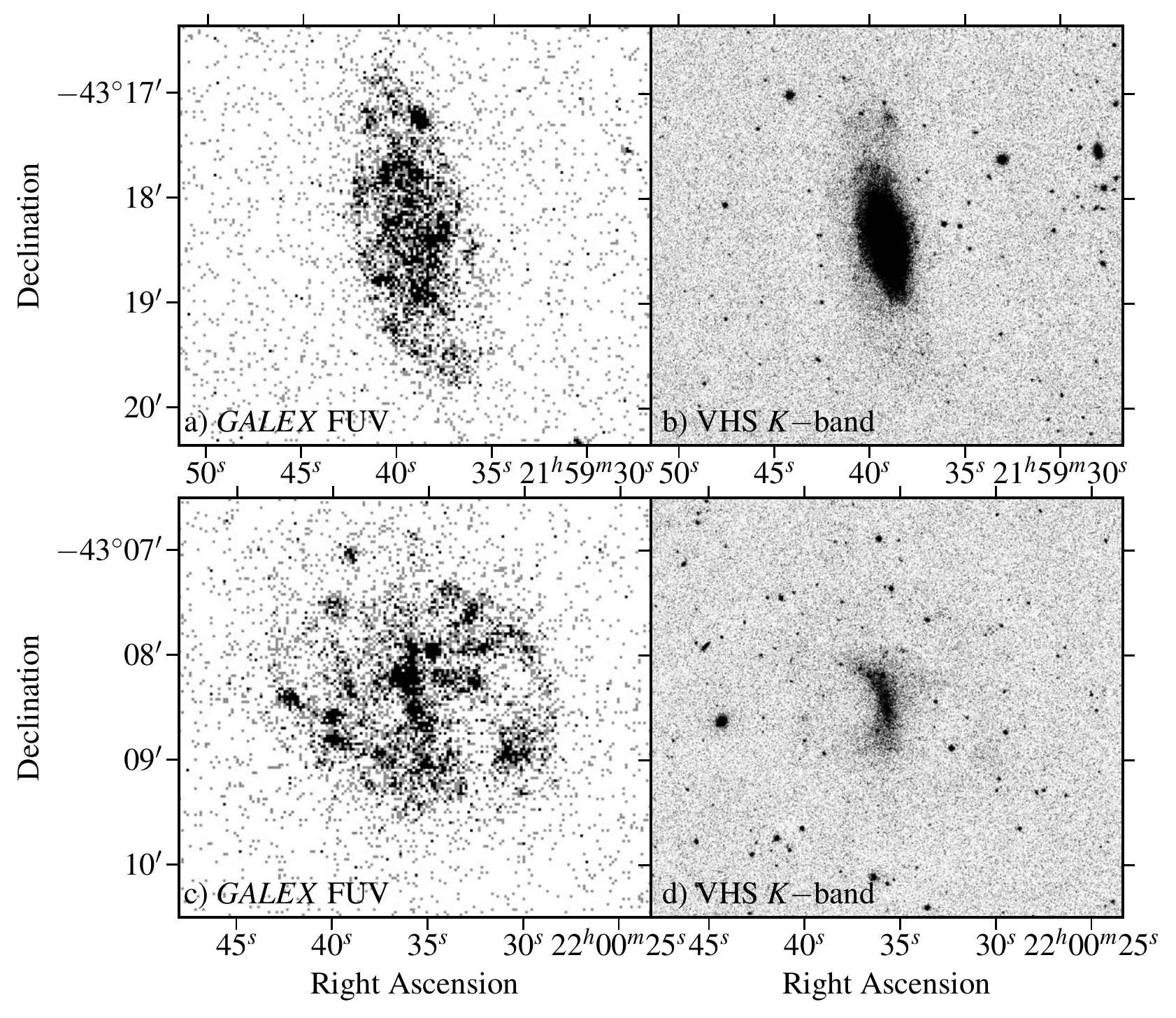}
	\caption{\textit{GALEX} FUV and VHS $K-$band images of NGC\,7162 (panels a and b) and NGC\,7162A (panels c and d) illustrating that both galaxies appear much more symmetric in the UV than the IR. We also show that NGC\,7162A is a Type 2 XUV galaxy with the emission from the FUV disk extending to much larger radii than the emission in the $K-$band.}
	\label{fig:ngc7162_a_vhs_galex}
\end{figure}

\begin{enumerate}[I]
	\item We start with the most promising candidate, NGC\,7162A, which appears to have a small twist in position angle between velocity field contours at the galaxy's centre compared with the outer regions (Fig.~\ref{fig:all_maps4} row b, centre column). We also see this in the derived position angle from the tilted ring fitting to the the full disk. However there is large uncertainty when looking at either the approaching or receding sides separately and the position angle is approximately constant across all radii within the errors (Fig.~\ref{fig:both_galaxies_kin15}e). The position angles from the 25\,mag\,arcsec$^{-2}$ $B-$band image and the value derived from the H\,\textsc{i} gas kinematics from SoFiA are offset by $\sim40^{\circ}$, indicating the gas does not rotate in accordance with the galaxy's optical major axis. This, along with the possible small twist in the position angle with radius, could indicate that an interaction occurred in the group's past, either sufficiently long ago that there are no other obvious signs of the interaction or that we simply are unable to detect any morphological signs due to the imaging limitations of our observations. 

    In this group, NGC\,7162A is the most H\,\textsc{i} rich galaxy with a H\,\textsc{i} mass excess of $\sim0.95$\,dex compared with isolated galaxies. This excess is surprising due to the presence of neighbouring galaxies with projected distances of $\sim126$\,kpc (NGC\,7162 and NGC\,7166), placing NGC\,7162A in the densest part of the group, where galaxies are more commonly H\,\textsc{i} deficient. The excess gas is unlikely to have been removed from NGC\,7162, which is also H\,\textsc{i} rich. However we are unable to identify an origin of the excess gas as we cannot detect any faint extra-planar gas inflowing onto NGC\,7162A from either the ASKAP-12 or ATCA data (Fig.~\ref{fig:all_maps4} left column, row b). Although NGC\,7162A is H\,\textsc{i} rich, it is normal with regards to its H\,\textsc{i} to stellar mass ratio, $\log(M_{\mathrm{HI}}/M_*)$, and specific star formation rate, $\log(\mathrm{sSFR})$, as it follows the scaling relations in figure~5 from \cite{Catinella2018} for $\log(M_{\mathrm{HI}}/M_*)$ vs $\log(M_*/\mathrm{M}_{\sun})$ and $\log(\mathrm{sSFR})$ (Table~\ref{table:galaxy_params}). The gas fraction of NGC\,7162A also agrees with the findings of \cite{Janowiecki2017} where group centrals have higher $\log(M_{\mathrm{HI}}/M_*)$ at fixed $\log(M_*/\mathrm{M}_{\sun})$. NGC\,7162A has a slightly lower star formation efficiency ($\mathrm{SFE}=\mathrm{SFR}/M_{\mathrm{HI}}$) of $\log(\mathrm{SFE}/\mathrm{yr})=-10.2$ compared with previous samples \citep[e.g. $\log(\mathrm{SFE}/\mathrm{yr})=-9.5,\,-9.95,\,-9.65$,][respectively]{Schiminovich2010,Huang2012,Wong2016}, which could be due to it having a higher disk stability against star formation.
    
	While we do not observe H\,\textsc{i} asymmetries in NGC\,7162A in either ASKAP or ATCA column density maps (Fig.~\ref{fig:all_maps4} row 2, left column), it does appear asymmetric in \textit{Spitzer} IRAC images and has asymmetry parameter values of 0.42 and 0.54 at 3.6\,$\mu$m and 4.5\,$\mu$m, respectively \citep[][]{Holwerda2014}. The asymmetry parameter is a quantitative measure of the asymmetry of a galaxy's brightness distribution with larger values indicating higher level of asymmetry. However, NGC\,7162A is a Type 2 extended ultraviolet disk (XUV-disk) galaxy and appears clumpy, but symmetric in \textit{GALEX} far-ultraviolet (FUV) images, demonstrating the need for multiple wavelengths to build up a more complete picture of the galaxy. Our classification of NGC\,7162A as a Type 2 XUV-disk explains the lack of any identifiable near-infrared features in the outer disk by \cite{Laine2014}. We can see this in Fig.~\ref{fig:ngc7162_a_vhs_galex}c and d, where \textit{GALEX} FUV emission extends to significantly larger radii compared with the VHS $K-$band emission concentrated in the galaxy's centre \citep[for details on XUV-disk galaxies, see e.g.][]{Thilker2007}. The star forming disk of XUV-disk galaxies is much more extended than the older stellar population and questions remain on how these galaxies form (e.g. do tidal interactions provide the gas reservoir with a large distribution across the galaxy's disk?). WALLABY will resolve the H\,\textsc{i} in XUV-disk galaxies across the southern sky in a variety of environments and will be able to shed light on the relation between XUV-disk and H\,\textsc{i} gas morphology. 

	\item NGC\,7162 is H\,\textsc{i} rich by $\sim0.72$\,dex. Similarly to NGC\,7162A, NGC\,7162 follows the scaling relations in figure~5 from \cite{Catinella2018} for $\log(M_{\mathrm{HI}}/M_*)$ vs $\log(M_*)$ and $\log(\mathrm{sSFR})$ (Table~\ref{table:galaxy_params}). We find the inclination angle decreases towards the outer edge of the disk, starting at $\sim70^{\circ}$ and decreasing to $\sim55^{\circ}$ at the largest measured radius. Unlike NGC\,7162A, NGC\,7162 has good agreement between the position angles from the 25\,mag\,arcsec$^{-2}$ $B-$band image with the derived value from the H\,\textsc{i} gas kinematics from SoFiA ($10^{\circ}$ vs $14^{\circ}$, respectively). Although NGC\,7162 has high asymmetry parameter values of 0.74 and 0.51 from \textit{Spitzer} IRAC 3.6\,$\mu$m and 4.5\,$\mu$m images, respectively \citep{Holwerda2014}, in GALEX FUV images it also appears symmetrical. NGC\,7162 may also be an Type 2 XUV-disk galaxy, however its inclination is too high to say with certainty.
    
	Referring to the ATCA column density map, we see that NGC\,7162 does have a small amount of extended H\,\textsc{i} emission to the north east, pointing in the direction of NGC\,7162A (Fig.~\ref{fig:all_maps4} row 1, left column, thick grey contour). This shows that NGC\,7162 has true extended emission, which we do not detect in the ASKAP-12 data because it is hidden in the residual sidelobes. The location of the extended emission and the lack of any extended emission detected around NGC\,7162A suggests that this could be either very early stages of an interaction within the group or the end of the gas accretion onto NGC\,7162 given its large H\,\textsc{i} excess. The extended emission is fairly faint and only accounts for $\sim2\%$ of the galaxy's total H\,\textsc{i} mass.

	\item ESO\,288-G025 has a H\,\textsc{i} excess of $\sim0.39$\,dex, which we consider to be H\,\textsc{i} normal (see Section~\ref{sec:hi_gas}). This is expected as ESO\,288-G025 has the largest separation from the group (e.g. projected distance of $\sim305$\,kpc to NGC\,7162). There is excellent agreement between the position angles from the optical $B-$band image with the derived value from the H\,\textsc{i} gas kinematics from SoFiA ($53^{\circ}$ vs $52^{\circ}$, respectively).

	\item ESO\,288-G033 is also H\,\textsc{i} rich (excess $\sim0.66$\,dex). There is a small offset ($10^{\circ}$) between the position angles from the optical $B-$band image with the derived value from the H\,\textsc{i} gas kinematics from SoFiA. The ATCA column density map of ESO\,288-G033 appears asymmetric with the H\,\textsc{i} emission more extended to the north (Fig.~\ref{fig:all_maps4} row 4, left column, thick grey contour). However, this may be a result of the increased noise around ESO\,288-G033, as it lies near the edge of the ATCA beam and we do not draw any conclusions on whether this indicates ESO\,288-G033 is interacting.
    
	\item We are unable to comment on internal signs of interactions in the dwarfs AM\,2159-434 and J220338-431131 due to their small size, low signal-to-noise and lack of ancillary data for comparison.
\end{enumerate}

The NGC\,7162 group has a single X-ray luminosity upper limit for NGC\,7166 of $\log(L_{\mathrm{X}})<40.65$\,erg\,s$^{-1}$ \citep{Beuing1999} and is not defined as X-ray luminous. Previous studies of X-ray luminosity and H\,\textsc{i} deficiency found X-ray luminous groups tend to be more H\,\textsc{i} deficient relative to groups without X-ray emission \citep{Chamaraux2004,Sengupta2006,Kilborn2009}. Hence, the lack of X-ray emission from the NGC\,7162 group is expected given the H\,\textsc{i} richness of the group, as all group spirals have H\,\textsc{i} excesses, with no galaxies detected to be deficient in H\,\textsc{i}.

All four spiral galaxies for which we can calculate dynamical or dark matter masses are dark matter dominated with dark matter fractions ranging from $f_{\mathrm{DM}}=0.81-0.95$. Although we are unable to use envelope tracing or tilted ring modelling for either dwarf galaxy, we attempt to derive approximate dynamical mass estimates up to a factor of $\sin^2(i)$, leaving the galaxy inclination angle as a free parameter. For AM\,2159-434, we derive a dynamical mass estimate of $\log(M_{\mathrm{dyn}} \times \sin^2(i)/\mathrm{M}_{\sun})\sim9.1\pm0.4$ (where we assume the rotational velocity to be half the velocity width, $w_{50}$, of the integrated spectrum, $\sim27$\,$\mathrm{km}\,\mathrm{s}^{-1}$, and the radius to be the radius in the moment 0 map, $\sim8.4$\,$\mathrm{kpc}$). Similarly we derive an approximate dynamical mass estimate for J220338-431131 of $\log(M_{\mathrm{dyn}} \times \sin^2(i)/\mathrm{M}_{\sun})\sim8.9\pm0.4$ (assumed rotational velocity $\sim18$\,$\mathrm{km}\,\mathrm{s}^{-1}$ and radius $\sim10.9$\,$\mathrm{kpc}$). As we mentioned, we are limited in our mass modelling to estimating the total dark matter content rather than differentiating among different dark matter density profiles. However, we still demonstrate the potential of ASKAP and WALLABY to provide dynamical and dark matter mass estimates for all HIPASS detected galaxies resolved by the ASKAP synthesised beam and map the dark matter distribution in gas-rich galaxies in the local Universe over the entire southern sky.

% -------------------------------------------------------------------------------------------------------------------------------------------------------------------------
\section{SUMMARY}
\label{sec:conclusion}

In this work we have presented first results from WALLABY early science observations of the NGC\,7162 galaxy group taken with ASKAP. We detected six galaxies in H\,\textsc{i}: NGC\,7162, NGC\,7162A, ESO\,288-G025, ESO\,288-G033, AM\,2159-434 and J220338-431131. ESO\,288-G033, AM\,2159-434 and J220338-431131 are newly detected group members. Additionally, these are the first H\,\textsc{i} detections and distance measurements of the dwarf galaxies AM\,2159-434 and J220338-431131. 

Using archival HIPASS and ATCA observations, we performed validation checks on the ASKAP-12 observations. Minor calibration errors and the incomplete uv-coverage of ASKAP-12 result in residual sidelobes in the final image cube. This leads to a loss of flux in the brightest sources in the ASKAP observations compared with the archival ATCA and HIPASS data. Full ASKAP (36 antennas) will not be subject to these challenges. In the ASKAP-12 data we are unable to detect any morphological signs of interactions, although we do see extended H\,\textsc{i} emission to the north from NGC\,7162 in the archival ATCA image cube. We are unable to see this H\,\textsc{i} extension in the ASKAP-12 image cube as it is either lost in the sidelobes present in the final data cube or resolved out by ASKAP. ASKAP also resolves out $\sim40\%$ of the diffuse H\,\textsc{i} emission in NGC\,7162A compared with HIPASS. Due to the H\,\textsc{i} richness of the spiral galaxies and the lack of significant indications of group interactions, it is likely that these galaxies are infalling for the first time and are yet to undergo any significant tidal interactions.

These ASKAP images have provided improved spatial and spectral resolution compared with HIPASS and ATCA observations and significantly increased field of view. The improved resolution and more circular beam compared with ATCA allow us to derive rotation curves and hence dynamical masses for all four spiral galaxies, and dark matter masses for NGC\,7162 and NGC\,7162A. We also attempt to derive rough estimates for the dynamical masses for the two dwarf galaxies. All observed spiral galaxies are dark matter dominated, with dark matter fractions $\sim0.81-0.95$. However, we did not have sufficient resolution of the central region of the galaxies to test different dark matter models.

This work demonstrates the power of full WALLABY to estimate the dark matter content of gas-rich galaxies across the southern sky. The large field of view of ASKAP made it possible to observe the entire NGC\,7162 group in a single observation, while the centre of the observed 30 square degree field was focused on a different target, the NGC\,7232 triplet (separated by $\sim3.8\degr$). Although the ASKAP-12 observations were carried out in $\sim175$\,h over 16 nights, full WALLABY will be able to achieve better sensitivity in a single 12\,h observation, using 36 antennas, and study resolved galaxy groups at different stages of evolution across the entire southern sky. The mosaicked image cube produced with ASKAP\textsc{soft} is publicly available and can be downloaded at \url{https://doi.org/102.100.100/74066}.

\section*{Acknowledgements}

This research was conducted by the Australian Research Council Centre of Excellence for All Sky Astrophysics in 3 Dimensions (ASTRO 3D), through project number CE170100013. This research has made use of the NASA/IPAC Extragalactic Database (NED) which is operated by the Jet Propulsion Laboratory, California Institute of Technology, under contract with the National Aeronautics and Space Administration. The Australia Telescope Compact Array and Parkes radio telescope are part of the Australia Telescope National Facility which is funded by the Australian Government for operation as a National Facility managed by CSIRO. This paper includes archived data obtained through the Australia Telescope Online Archive (\url{http://atoa.atnf.csiro.au}). The Australian SKA Pathfinder is part of the Australia Telescope National Facility which is managed by CSIRO. Operation of ASKAP is funded by the Australian Government with support from the National Collaborative Research Infrastructure Strategy. ASKAP uses the resources of the Pawsey Supercomputing Centre. Establishment of ASKAP, the Murchison Radio-astronomy Observatory (MRO) and the Pawsey Supercomputing Centre are initiatives of the Australian Government, with support from the Government of Western Australia and the Science and Industry Endowment Fund. We acknowledge the Wajarri Yamatji as the traditional owners of the Observatory site. We also thank the MRO site staff. 

This work is based in part on observations made with the Galaxy Evolution Explorer (GALEX). GALEX is a NASA Small Explorer, whose mission was developed in cooperation with the Centre National d'Etudes Spatiales (CNES) of France and the Korean Ministry of Science and Technology. GALEX is operated for NASA by the California Institute of Technology under NASA contract NAS5-98034.

This project used public archival data from the Dark Energy Survey (DES). Funding for the DES Projects has been provided by the U.S. Department of Energy, the U.S. National Science Foundation, the Ministry of Science and Education of Spain, the Science and Technology Facilities Council of the United Kingdom, the Higher Education Funding Council for England, the National Center for Supercomputing Applications at the University of Illinois at Urbana-Champaign, the Kavli Institute of Cosmological Physics at the University of Chicago, the Center for Cosmology and Astro-Particle Physics at the Ohio State University, the Mitchell Institute for Fundamental Physics and Astronomy at Texas A\&M University, Financiadora de Estudos e Projetos, Funda{\c c}{\~a}o Carlos Chagas Filho de Amparo {\`a} Pesquisa do Estado do Rio de Janeiro, Conselho Nacional de Desenvolvimento Cient{\'i}fico e Tecnol{\'o}gico and the Minist{\'e}rio da Ci{\^e}ncia, Tecnologia e Inova{\c c}{\~a}o, the Deutsche Forschungs\-gemeinschaft, and the Collaborating Institutions in the Dark Energy Survey.

The Collaborating Institutions are Argonne National Laboratory, the University of California at Santa Cruz, the University of Cambridge, Centro de Investigaciones Energ{\'e}ticas, Medioambientales y Tecnol{\'o}gicas-Madrid, the University of Chicago, University College London, the DES-Brazil Consortium, the University of Edinburgh, the Eid\-gen{\"o}ssische Technische Hochschule (ETH) Z{\"u}rich,  Fermi National Accelerator Laboratory, the University of Illinois at Urbana-Champaign, the Institut de Ci{\`e}ncies de l'Espai (IEEC/CSIC), the Institut de F{\'i}sica d'Altes Energies, Lawrence Berkeley National Laboratory, the Ludwig-Maximilians Universit{\"a}t M{\"u}nchen and the associated Excellence Cluster Universe, the University of Michigan, the National Optical Astronomy Observatory, the University of Nottingham, The Ohio State University, the OzDES Membership Consortium, the University of Pennsylvania, the University of Portsmouth, SLAC National Accelerator Laboratory, Stanford University, the University of Sussex, and Texas A\&M University.

Based in part on observations at Cerro Tololo Inter-American Observatory, National Optical Astronomy Observatory, which is operated by the Association of Universities for Research in Astronomy (AURA) under a cooperative agreement with the National Science Foundation.

%%%%%%%%%%%%%%%%%%%%%%%%%%%%%%%%%%%%%%%%%%%%%%%%%%



\bibliographystyle{mnras}
\bibliography{master}

\begin{thebibliography}{}
\makeatletter
\relax
\def\mn@urlcharsother{\let\do\@makeother \do\$\do\&\do\#\do\^\do\_\do\%\do\~}
\def\mn@doi{\begingroup\mn@urlcharsother \@ifnextchar [ {\mn@doi@}
  {\mn@doi@[]}}
\def\mn@doi@[#1]#2{\def\@tempa{#1}\ifx\@tempa\@empty \href
  {http://dx.doi.org/#2} {doi:#2}\else \href {http://dx.doi.org/#2} {#1}\fi
  \endgroup}
\def\mn@eprint#1#2{\mn@eprint@#1:#2::\@nil}
\def\mn@eprint@arXiv#1{\href {http://arxiv.org/abs/#1} {{\tt arXiv:#1}}}
\def\mn@eprint@dblp#1{\href {http://dblp.uni-trier.de/rec/bibtex/#1.xml}
  {dblp:#1}}
\def\mn@eprint@#1:#2:#3:#4\@nil{\def\@tempa {#1}\def\@tempb {#2}\def\@tempc
  {#3}\ifx \@tempc \@empty \let \@tempc \@tempb \let \@tempb \@tempa \fi \ifx
  \@tempb \@empty \def\@tempb {arXiv}\fi \@ifundefined
  {mn@eprint@\@tempb}{\@tempb:\@tempc}{\expandafter \expandafter \csname
  mn@eprint@\@tempb\endcsname \expandafter{\@tempc}}}

\bibitem[\protect\citeauthoryear{{Abbott} et~al.,}{{Abbott}
  et~al.}{2018}]{Abbott2018}
{Abbott} T.~M.~C.,  et~al., 2018, preprint, \href
  {http://adsabs.harvard.edu/abs/2018arXiv180103181A} {} (\mn@eprint {arXiv}
  {1801.03181})

\bibitem[\protect\citeauthoryear{{Applebaum}}{{Applebaum}}{1976}]{Applebaum1976}
{Applebaum} S.~P.,  1976, \mn@doi [IEEE Transactions on Antennas and
  Propagation] {10.1109/TAP.1976.1141417}, \href
  {http://adsabs.harvard.edu/abs/1976ITAP...24..585A} {24, 585}

\bibitem[\protect\citeauthoryear{{Barnes} et~al.,}{{Barnes}
  et~al.}{2001}]{Barnes2001}
{Barnes} D.~G.,  et~al., 2001, \mn@doi [\mnras]
  {10.1046/j.1365-8711.2001.04102.x}, \href
  {http://adsabs.harvard.edu/abs/2001MNRAS.322..486B} {322, 486}

\bibitem[\protect\citeauthoryear{{Begeman}, {Broeils}  \& {Sanders}}{{Begeman}
  et~al.}{1991}]{Begeman1991}
{Begeman} K.~G.,  {Broeils} A.~H.,   {Sanders} R.~H.,  1991, \mn@doi [\mnras]
  {10.1093/mnras/249.3.523}, \href
  {http://adsabs.harvard.edu/abs/1991MNRAS.249..523B} {249, 523}

\bibitem[\protect\citeauthoryear{{Beuing}, {Dobereiner}, {Bohringer}  \&
  {Bender}}{{Beuing} et~al.}{1999}]{Beuing1999}
{Beuing} J.,  {Dobereiner} S.,  {Bohringer} H.,   {Bender} R.,  1999, \mn@doi
  [\mnras] {10.1046/j.1365-8711.1999.02108.x}, \href
  {http://adsabs.harvard.edu/abs/1999MNRAS.302..209B} {302, 209}

\bibitem[\protect\citeauthoryear{Boselli \& Gavazzi}{Boselli \&
  Gavazzi}{2009}]{Boselli2009}
Boselli A.,  Gavazzi G.,  2009, \mn@doi [aap] {10.1051/0004-6361/200912658},
  508, 201

\bibitem[\protect\citeauthoryear{Brough, Forbes, Kilborn  \& Couch}{Brough
  et~al.}{2006}]{Brough2006}
Brough S.,  Forbes D.~A.,  Kilborn V.~A.,   Couch W.,  2006, \mn@doi [Monthly
  Notices of the Royal Astronomical Society]
  {10.1111/j.1365-2966.2006.10542.x}, 370, 1223

\bibitem[\protect\citeauthoryear{{Catinella} et~al.,}{{Catinella}
  et~al.}{2018}]{Catinella2018}
{Catinella} B.,  et~al., 2018, \mn@doi [\mnras] {10.1093/mnras/sty089}, \href
  {https://ui.adsabs.harvard.edu/#abs/2018MNRAS.476..875C} {476, 875}

\bibitem[\protect\citeauthoryear{{Chabrier}}{{Chabrier}}{2003}]{Chabrier2003}
{Chabrier} G.,  2003, \mn@doi [\pasp] {10.1086/376392}, \href
  {http://adsabs.harvard.edu/abs/2003PASP..115..763C} {115, 763}

\bibitem[\protect\citeauthoryear{{Chamaraux} \& {Masnou}}{{Chamaraux} \&
  {Masnou}}{2004}]{Chamaraux2004}
{Chamaraux} P.,  {Masnou} J.-L.,  2004, \mn@doi [\mnras]
  {10.1111/j.1365-2966.2004.07226.x}, \href
  {http://adsabs.harvard.edu/abs/2004MNRAS.347..541C} {347, 541}

\bibitem[\protect\citeauthoryear{{Chippendale} et~al.,}{{Chippendale}
  et~al.}{2015}]{Chippendale2015}
{Chippendale} A.~P.,  et~al., 2015, in 2015 International Conference on
  Electromagnetics in Advanced Applications (ICEAA), p. 541-544. pp 541--544
  (\mn@eprint {arXiv} {1509.00544}), \mn@doi{10.1109/ICEAA.2015.7297174}

\bibitem[\protect\citeauthoryear{{Chung}, {van Gorkom}, {Kenney}, {Crowl}  \&
  {Vollmer}}{{Chung} et~al.}{2009}]{Chung2009}
{Chung} A.,  {van Gorkom} J.~H.,  {Kenney} J.~D.~P.,  {Crowl} H.,   {Vollmer}
  B.,  2009, \mn@doi [\aj] {10.1088/0004-6256/138/6/1741}, \href
  {http://adsabs.harvard.edu/abs/2009AJ....138.1741C} {138, 1741}

\bibitem[\protect\citeauthoryear{{Cluver}, {Jarrett}, {Dale}, {Smith}, {August}
   \& {Brown}}{{Cluver} et~al.}{2017}]{Cluver2017}
{Cluver} M.~E.,  {Jarrett} T.~H.,  {Dale} D.~A.,  {Smith} J.-D.~T.,  {August}
  T.,   {Brown} M.~J.~I.,  2017, \mn@doi [\apj] {10.3847/1538-4357/aa92c7},
  \href {http://adsabs.harvard.edu/abs/2017ApJ...850...68C} {850, 68}

\bibitem[\protect\citeauthoryear{{Cornwell}}{{Cornwell}}{2008}]{Cornwell2008a}
{Cornwell} T.~J.,  2008, \mn@doi [IEEE Journal of Selected Topics in Signal
  Processing] {10.1109/JSTSP.2008.2006388}, \href
  {http://adsabs.harvard.edu/abs/2008ISTSP...2..793C} {2, 793}

\bibitem[\protect\citeauthoryear{{Cornwell}, {Golap}  \&
  {Bhatnagar}}{{Cornwell} et~al.}{2008}]{Cornwell2008b}
{Cornwell} T.~J.,  {Golap} K.,   {Bhatnagar} S.,  2008, \mn@doi [IEEE Journal
  of Selected Topics in Signal Processing] {10.1109/JSTSP.2008.2005290}, \href
  {http://adsabs.harvard.edu/abs/2008ISTSP...2..647C} {2, 647}

\bibitem[\protect\citeauthoryear{Cortese, Catinella, Boissier, Boselli  \&
  Heinis}{Cortese et~al.}{2011}]{Cortese2011}
Cortese L.,  Catinella B.,  Boissier S.,  Boselli A.,   Heinis S.,  2011,
  \mn@doi [\mnras] {10.1111/j.1365-2966.2011.18822.x}, 415, 1797

\bibitem[\protect\citeauthoryear{{DeBoer} et~al.,}{{DeBoer}
  et~al.}{2009}]{DeBoer2009}
{DeBoer} D.~R.,  et~al., 2009, \mn@doi [IEEE Proceedings]
  {10.1109/JPROC.2009.2016516}, \href
  {http://adsabs.harvard.edu/abs/2009IEEEP..97.1507D} {97, 1507}

\bibitem[\protect\citeauthoryear{{D{\'e}nes}, {Kilborn}  \&
  {Koribalski}}{{D{\'e}nes} et~al.}{2014}]{Denes2014}
{D{\'e}nes} H.,  {Kilborn} V.~A.,   {Koribalski} B.~S.,  2014, \mn@doi [\mnras]
  {10.1093/mnras/stu1337}, \href
  {http://adsabs.harvard.edu/abs/2014MNRAS.444..667D} {444, 667}

\bibitem[\protect\citeauthoryear{{Dressler}}{{Dressler}}{1980}]{Dressler1980}
{Dressler} A.,  1980, \mn@doi [\apj] {10.1086/157753}, \href
  {http://adsabs.harvard.edu/abs/1980ApJ...236..351D} {236, 351}

\bibitem[\protect\citeauthoryear{{Driver} et~al.,}{{Driver}
  et~al.}{2011}]{Driver2011}
{Driver} S.~P.,  et~al., 2011, \mn@doi [\mnras]
  {10.1111/j.1365-2966.2010.18188.x}, \href
  {http://adsabs.harvard.edu/abs/2011MNRAS.413..971D} {413, 971}

\bibitem[\protect\citeauthoryear{{Duffy}, {Meyer}, {Staveley-Smith}, {Bernyk},
  {Croton}, {Koribalski}, {Gerstmann}  \& {Westerlund}}{{Duffy}
  et~al.}{2012}]{Duffy2012}
{Duffy} A.~R.,  {Meyer} M.~J.,  {Staveley-Smith} L.,  {Bernyk} M.,  {Croton}
  D.~J.,  {Koribalski} B.~S.,  {Gerstmann} D.,   {Westerlund} S.,  2012,
  \mn@doi [\mnras] {10.1111/j.1365-2966.2012.21987.x}, \href
  {http://adsabs.harvard.edu/abs/2012MNRAS.426.3385D} {426, 3385}

\bibitem[\protect\citeauthoryear{{English}, {Koribalski}, {Bland-Hawthorn},
  {Freeman}  \& {McCain}}{{English} et~al.}{2010}]{English2010}
{English} J.,  {Koribalski} B.,  {Bland-Hawthorn} J.,  {Freeman} K.~C.,
  {McCain} C.~F.,  2010, \mn@doi [\aj] {10.1088/0004-6256/139/1/102}, \href
  {http://adsabs.harvard.edu/abs/2010AJ....139..102E} {139, 102}

\bibitem[\protect\citeauthoryear{{Fouque}, {Gourgoulhon}, {Chamaraux}  \&
  {Paturel}}{{Fouque} et~al.}{1992}]{Fouque1992}
{Fouque} P.,  {Gourgoulhon} E.,  {Chamaraux} P.,   {Paturel} G.,  1992, \aaps,
  \href {http://adsabs.harvard.edu/abs/1992A%26AS...93..211F} {93, 211}

\bibitem[\protect\citeauthoryear{{Freeland}, {Stilp}  \& {Wilcots}}{{Freeland}
  et~al.}{2009}]{Freeland2009}
{Freeland} E.,  {Stilp} A.,   {Wilcots} E.,  2009, \mn@doi [\aj]
  {10.1088/0004-6256/138/1/295}, \href
  {http://adsabs.harvard.edu/abs/2009AJ....138..295F} {138, 295}

\bibitem[\protect\citeauthoryear{{Giovanelli} \& {Haynes}}{{Giovanelli} \&
  {Haynes}}{1985}]{Giovanelli1985}
{Giovanelli} R.,  {Haynes} M.~P.,  1985, \mn@doi [\apj] {10.1086/163170}, \href
  {http://adsabs.harvard.edu/abs/1985ApJ...292..404G} {292, 404}

\bibitem[\protect\citeauthoryear{{Gourgoulhon}, {Chamaraux}  \&
  {Fouque}}{{Gourgoulhon} et~al.}{1992}]{Gourgoulhon1992}
{Gourgoulhon} E.,  {Chamaraux} P.,   {Fouque} P.,  1992, \aap, \href
  {http://adsabs.harvard.edu/abs/1992A%26A...255...69G} {255, 69}

\bibitem[\protect\citeauthoryear{{Gunn} \& {Gott}}{{Gunn} \&
  {Gott}}{1972}]{Gunn1972}
{Gunn} J.~E.,  {Gott} III J.~R.,  1972, \mn@doi [\apj] {10.1086/151605}, \href
  {http://adsabs.harvard.edu/abs/1972ApJ...176....1G} {176, 1}

\bibitem[\protect\citeauthoryear{Hampson et~al.,}{Hampson
  et~al.}{2012}]{Hampson2012}
Hampson G.,  et~al., 2012, in Electromagnetics in Advanced Applications
  (ICEAA), 2012 International Conference on. pp 807--809,
  \mn@doi{10.1109/ICEAA.2012.6328742}

\bibitem[\protect\citeauthoryear{Hay \& O'Sullivan}{Hay \&
  O'Sullivan}{2008}]{Hay2008}
Hay S.,  O'Sullivan J.,  2008, \mn@doi [Radio Science] {10.1029/2007RS003798},
  43

\bibitem[\protect\citeauthoryear{Haynes \& Giovanelli}{Haynes \&
  Giovanelli}{1984}]{Haynes1984}
Haynes M.~P.,  Giovanelli R.,  1984, The Astronomical Journal, 89

\bibitem[\protect\citeauthoryear{{Haynes} et~al.,}{{Haynes}
  et~al.}{2018}]{Haynes2018}
{Haynes} M.~P.,  et~al., 2018, preprint, \href
  {http://adsabs.harvard.edu/abs/2018arXiv180511499H} {} (\mn@eprint {arXiv}
  {1805.11499})

\bibitem[\protect\citeauthoryear{{Heald} et~al.,}{{Heald}
  et~al.}{2011}]{Heald2011}
{Heald} G.,  et~al., 2011, \mn@doi [\aap] {10.1051/0004-6361/201015938}, \href
  {http://adsabs.harvard.edu/abs/2011A%26A...526A.118H} {526, A118}

\bibitem[\protect\citeauthoryear{Hess \& Wilcots}{Hess \&
  Wilcots}{2013}]{Hess2013}
Hess K.~M.,  Wilcots E.~M.,  2013, \mn@doi [Astronomical Journal]
  {10.1088/0004-6256/146/5/124}, 146

\bibitem[\protect\citeauthoryear{Hess, Cluver, Yahya, Leisman, Serra, Lucero,
  Passmoor  \& Carignan}{Hess et~al.}{2017}]{Hess2017}
Hess K.~M.,  Cluver M.~E.,  Yahya S.,  Leisman L.,  Serra P.,  Lucero D.~M.,
  Passmoor S.~S.,   Carignan C.,  2017, \mn@doi [Monthly Notices of the Royal
  Astronomical Society] {10.1093/mnras/stw2338}, 464, 957

\bibitem[\protect\citeauthoryear{{Heywood} et~al.,}{{Heywood}
  et~al.}{2016}]{Heywood2016}
{Heywood} I.,  et~al., 2016, \mn@doi [\mnras] {10.1093/mnras/stw186}, \href
  {http://adsabs.harvard.edu/abs/2016MNRAS.457.4160H} {457, 4160}

\bibitem[\protect\citeauthoryear{{Holwerda} et~al.,}{{Holwerda}
  et~al.}{2014}]{Holwerda2014}
{Holwerda} B.~W.,  et~al., 2014, \mn@doi [\apj] {10.1088/0004-637X/781/1/12},
  \href {http://adsabs.harvard.edu/abs/2014ApJ...781...12H} {781, 12}

\bibitem[\protect\citeauthoryear{{Hotan} et~al.,}{{Hotan}
  et~al.}{2014}]{Hotan2014}
{Hotan} A.~W.,  et~al., 2014, \mn@doi [\pasa] {10.1017/pasa.2014.36}, \href
  {http://adsabs.harvard.edu/abs/2014PASA...31...41H} {31, e041}

\bibitem[\protect\citeauthoryear{{Huang}, {Haynes}, {Giovanelli}  \&
  {Brinchmann}}{{Huang} et~al.}{2012}]{Huang2012}
{Huang} S.,  {Haynes} M.~P.,  {Giovanelli} R.,   {Brinchmann} J.,  2012,
  \mn@doi [\apj] {10.1088/0004-637X/756/2/113}, \href
  {http://adsabs.harvard.edu/abs/2012ApJ...756..113H} {756, 113}

\bibitem[\protect\citeauthoryear{Jaff{\'{e}}, Smith, Candlish, Poggianti, Sheen
   \& Verheijen}{Jaff{\'{e}} et~al.}{2015}]{Jaffe2015}
Jaff{\'{e}} Y.~L.,  Smith R.,  Candlish G.~N.,  Poggianti B.~M.,  Sheen Y.~K.,
   Verheijen M.~A.,  2015, \mn@doi [\mnras] {10.1093/mnras/stv100}, 448

\bibitem[\protect\citeauthoryear{{Janowiecki}, {Catinella}, {Cortese},
  {Saintonge}, {Brown}  \& {Wang}}{{Janowiecki} et~al.}{2017}]{Janowiecki2017}
{Janowiecki} S.,  {Catinella} B.,  {Cortese} L.,  {Saintonge} A.,  {Brown} T.,
   {Wang} J.,  2017, \mn@doi [\mnras] {10.1093/mnras/stx046}, \href
  {http://adsabs.harvard.edu/abs/2017MNRAS.466.4795J} {466, 4795}

\bibitem[\protect\citeauthoryear{{Jones}, {Haynes}, {Giovanelli}  \&
  {Moorman}}{{Jones} et~al.}{2018}]{Jones2018}
{Jones} M.~G.,  {Haynes} M.~P.,  {Giovanelli} R.,   {Moorman} C.,  2018,
  \mn@doi [\mnras] {10.1093/mnras/sty521}, \href
  {http://adsabs.harvard.edu/abs/2018MNRAS.477....2J} {477, 2}

\bibitem[\protect\citeauthoryear{Kenney, van Gorkom  \& Vollmer}{Kenney
  et~al.}{2004}]{Kenney2004}
Kenney J. D.~P.,  van Gorkom J.~H.,   Vollmer B.,  2004, \mn@doi [Aj]
  {10.1086/420805}, 127, 3361

\bibitem[\protect\citeauthoryear{{Kern}, {Kilborn}, {Forbes}  \&
  {Koribalski}}{{Kern} et~al.}{2008}]{Kern2008}
{Kern} K.~M.,  {Kilborn} V.~A.,  {Forbes} D.~A.,   {Koribalski} B.,  2008,
  \mn@doi [\mnras] {10.1111/j.1365-2966.2007.12693.x}, \href
  {http://adsabs.harvard.edu/abs/2008MNRAS.384..305K} {384, 305}

\bibitem[\protect\citeauthoryear{Kilborn, Koribalski, Forbes, Barnes  \&
  Musgrave}{Kilborn et~al.}{2005}]{Kilborn2005}
Kilborn V.~A.,  Koribalski B.~S.,  Forbes D.~A.,  Barnes D.~G.,   Musgrave
  R.~C.,  2005, \mn@doi [Monthly Notices of the Royal Astronomical Society]
  {10.1111/j.1365-2966.2004.08450.x}, 356, 77

\bibitem[\protect\citeauthoryear{Kilborn, Forbes, Barnes, Koribalski, Brough
  \& Kern}{Kilborn et~al.}{2009}]{Kilborn2009}
Kilborn V.~A.,  Forbes D.~A.,  Barnes D.~G.,  Koribalski B.~S.,  Brough S.,
  Kern K.,  2009, \mn@doi [Monthly Notices of the Royal Astronomical Society]
  {10.1111/j.1365-2966.2009.15587.x}, 400, 1962

\bibitem[\protect\citeauthoryear{{Koribalski}}{{Koribalski}}{2012}]{Koribalski2012}
{Koribalski} B.~S.,  2012, \mn@doi [\pasa] {10.1071/AS12030}, \href
  {http://adsabs.harvard.edu/abs/2012PASA...29..359K} {29, 359}

\bibitem[\protect\citeauthoryear{Koribalski \& Dickey}{Koribalski \&
  Dickey}{2004}]{Koribalski2004}
Koribalski B.,  Dickey J.,  2004, \mn@doi [Monthly Notices of the Royal
  Astronomical Society] {10.1111/j.1365-2966.2004.07444.x}, 348, 1255

\bibitem[\protect\citeauthoryear{{Koribalski} \&
  {L{\'o}pez-S{\'a}nchez}}{{Koribalski} \&
  {L{\'o}pez-S{\'a}nchez}}{2009}]{Koribalski2009}
{Koribalski} B.~S.,  {L{\'o}pez-S{\'a}nchez} {\'A}.~R.,  2009, \mn@doi [\mnras]
  {10.1111/j.1365-2966.2009.15610.x}, \href
  {http://adsabs.harvard.edu/abs/2009MNRAS.400.1749K} {400, 1749}

\bibitem[\protect\citeauthoryear{Koribalski \& Manthey}{Koribalski \&
  Manthey}{2005}]{Koribalski2005}
Koribalski B.,  Manthey E.,  2005, \mn@doi [Monthly Notices of the Royal
  Astronomical Society] {10.1111/j.1365-2966.2005.08803.x}, 358, 202

\bibitem[\protect\citeauthoryear{{Koribalski} et~al.,}{{Koribalski}
  et~al.}{2018}]{Koribalski2018}
{Koribalski} B.~S.,  et~al., 2018, \mn@doi [\mnras] {10.1093/mnras/sty479},
  \href {http://adsabs.harvard.edu/abs/2018MNRAS.tmp..467K} {}

\bibitem[\protect\citeauthoryear{{Laine} et~al.,}{{Laine}
  et~al.}{2014}]{Laine2014}
{Laine} S.,  et~al., 2014, \mn@doi [\mnras] {10.1093/mnras/stu1642}, \href
  {http://adsabs.harvard.edu/abs/2014MNRAS.444.3015L} {444, 3015}

\bibitem[\protect\citeauthoryear{{Larson}, {Tinsley}  \& {Caldwell}}{{Larson}
  et~al.}{1980}]{Larson1980}
{Larson} R.~B.,  {Tinsley} B.~M.,   {Caldwell} C.~N.,  1980, \mn@doi [\apj]
  {10.1086/157917}, \href {http://adsabs.harvard.edu/abs/1980ApJ...237..692L}
  {237, 692}

\bibitem[\protect\citeauthoryear{{Lauberts} \& {Valentijn}}{{Lauberts} \&
  {Valentijn}}{1989}]{Lauberts1989}
{Lauberts} A.,  {Valentijn} E.~A.,  1989, {The surface photometry catalogue of
  the ESO-Uppsala galaxies}

\bibitem[\protect\citeauthoryear{{Longhetti} \& {Saracco}}{{Longhetti} \&
  {Saracco}}{2009}]{Longhetti2009}
{Longhetti} M.,  {Saracco} P.,  2009, \mn@doi [\mnras]
  {10.1111/j.1365-2966.2008.14375.x}, \href
  {http://adsabs.harvard.edu/abs/2009MNRAS.394..774L} {394, 774}

\bibitem[\protect\citeauthoryear{{Maia}, {da Costa}  \& {Latham}}{{Maia}
  et~al.}{1989}]{Maia1989}
{Maia} M.~A.~G.,  {da Costa} L.~N.,   {Latham} D.~W.,  1989, \mn@doi [\apjs]
  {10.1086/191328}, \href {http://adsabs.harvard.edu/abs/1989ApJS...69..809M}
  {69, 809}

\bibitem[\protect\citeauthoryear{{McConnell} et~al.,}{{McConnell}
  et~al.}{2016}]{McConnell2016}
{McConnell} D.,  et~al., 2016, \mn@doi [\pasa] {10.1017/pasa.2016.37}, \href
  {http://adsabs.harvard.edu/abs/2016PASA...33...42M} {33, e042}

\bibitem[\protect\citeauthoryear{{McMahon}, {Banerji}, {Gonzalez}, {Koposov},
  {Bejar}, {Lodieu}, {Rebolo}  \& {VHS Collaboration}}{{McMahon}
  et~al.}{2013}]{McMahon2013}
{McMahon} R.~G.,  {Banerji} M.,  {Gonzalez} E.,  {Koposov} S.~E.,  {Bejar}
  V.~J.,  {Lodieu} N.,  {Rebolo} R.,   {VHS Collaboration} 2013, The Messenger,
  \href {http://adsabs.harvard.edu/abs/2013Msngr.154...35M} {154, 35}

\bibitem[\protect\citeauthoryear{{Meyer} et~al.,}{{Meyer}
  et~al.}{2004}]{Meyer2004}
{Meyer} M.~J.,  et~al., 2004, \mn@doi [\mnras]
  {10.1111/j.1365-2966.2004.07710.x}, \href
  {http://adsabs.harvard.edu/abs/2004MNRAS.350.1195M} {350, 1195}

\bibitem[\protect\citeauthoryear{{Moore}, {Katz}, {Lake}, {Dressler}  \&
  {Oemler}}{{Moore} et~al.}{1996}]{Moore1996}
{Moore} B.,  {Katz} N.,  {Lake} G.,  {Dressler} A.,   {Oemler} A.,  1996,
  \mn@doi [\nat] {10.1038/379613a0}, \href
  {http://adsabs.harvard.edu/abs/1996Natur.379..613M} {379, 613}

\bibitem[\protect\citeauthoryear{{Moore}, {Lake}  \& {Katz}}{{Moore}
  et~al.}{1998}]{Moore1998}
{Moore} B.,  {Lake} G.,   {Katz} N.,  1998, \mn@doi [\apj] {10.1086/305264},
  \href {http://adsabs.harvard.edu/abs/1998ApJ...495..139M} {495, 139}

\bibitem[\protect\citeauthoryear{{Moore}, {Lake}, {Quinn}  \& {Stadel}}{{Moore}
  et~al.}{1999}]{Moore1999}
{Moore} B.,  {Lake} G.,  {Quinn} T.,   {Stadel} J.,  1999, \mn@doi [\mnras]
  {10.1046/j.1365-8711.1999.02345.x}, \href
  {http://adsabs.harvard.edu/abs/1999MNRAS.304..465M} {304, 465}

\bibitem[\protect\citeauthoryear{{Mu{\~n}oz-Mateos} et~al.,}{{Mu{\~n}oz-Mateos}
  et~al.}{2015}]{Munoz2015}
{Mu{\~n}oz-Mateos} J.~C.,  et~al., 2015, \mn@doi [\apjs]
  {10.1088/0067-0049/219/1/3}, \href
  {http://adsabs.harvard.edu/abs/2015ApJS..219....3M} {219, 3}

\bibitem[\protect\citeauthoryear{{Navarro}, {Frenk}  \& {White}}{{Navarro}
  et~al.}{1997}]{Navarro1997}
{Navarro} J.~F.,  {Frenk} C.~S.,   {White} S.~D.~M.,  1997, \mn@doi [\apj]
  {10.1086/304888}, \href {http://adsabs.harvard.edu/abs/1997ApJ...490..493N}
  {490, 493}

\bibitem[\protect\citeauthoryear{Odekon et~al.,}{Odekon
  et~al.}{2016}]{Odekon2016}
Odekon M.~C.,  et~al., 2016, \mn@doi [The Astrophysical Journal]
  {10.3847/0004-637X/824/2/110}, 824, 1

\bibitem[\protect\citeauthoryear{{Oh}, {de Blok}, {Walter}, {Brinks}  \&
  {Kennicutt}}{{Oh} et~al.}{2008}]{Oh2008}
{Oh} S.-H.,  {de Blok} W.~J.~G.,  {Walter} F.,  {Brinks} E.,   {Kennicutt} Jr.
  R.~C.,  2008, \mn@doi [\aj] {10.1088/0004-6256/136/6/2761}, \href
  {http://adsabs.harvard.edu/abs/2008AJ....136.2761O} {136, 2761}

\bibitem[\protect\citeauthoryear{{Oh}, {de Blok}, {Brinks}, {Walter}  \&
  {Kennicutt}}{{Oh} et~al.}{2011}]{Oh2011}
{Oh} S.-H.,  {de Blok} W.~J.~G.,  {Brinks} E.,  {Walter} F.,   {Kennicutt} Jr.
  R.~C.,  2011, \mn@doi [\aj] {10.1088/0004-6256/141/6/193}, \href
  {http://adsabs.harvard.edu/abs/2011AJ....141..193O} {141, 193}

\bibitem[\protect\citeauthoryear{{Oh} et~al.,}{{Oh} et~al.}{2015}]{Oh2015}
{Oh} S.-H.,  et~al., 2015, \mn@doi [\aj] {10.1088/0004-6256/149/6/180}, \href
  {http://adsabs.harvard.edu/abs/2015AJ....149..180O} {149, 180}

\bibitem[\protect\citeauthoryear{{Oh}, {Staveley-Smith}, {Spekkens}, {Kamphuis}
   \& {Koribalski}}{{Oh} et~al.}{2018}]{Oh2018}
{Oh} S.-H.,  {Staveley-Smith} L.,  {Spekkens} K.,  {Kamphuis} P.,
  {Koribalski} B.~S.,  2018, \mn@doi [\mnras] {10.1093/mnras/stx2304}, \href
  {http://adsabs.harvard.edu/abs/2018MNRAS.473.3256O} {473, 3256}

\bibitem[\protect\citeauthoryear{{Oosterloo}, {Morganti}, {Sadler}, {van der
  Hulst}  \& {Serra}}{{Oosterloo} et~al.}{2007}]{Oosterloo2007}
{Oosterloo} T.~A.,  {Morganti} R.,  {Sadler} E.~M.,  {van der Hulst} T.,
  {Serra} P.,  2007, \mn@doi [A\&A] {10.1051/0004-6361:20066384}, \href
  {http://adsabs.harvard.edu/abs/2007A%26A...465..787O} {465, 787}

\bibitem[\protect\citeauthoryear{{Pisano}, {Barnes}, {Staveley-Smith},
  {Gibson}, {Kilborn}  \& {Freeman}}{{Pisano} et~al.}{2011}]{Pisano2011}
{Pisano} D.~J.,  {Barnes} D.~G.,  {Staveley-Smith} L.,  {Gibson} B.~K.,
  {Kilborn} V.~A.,   {Freeman} K.~C.,  2011, \mn@doi [Astrophysical Journal,
  Supplement Series] {10.1088/0067-0049/197/2/28}, 197

\bibitem[\protect\citeauthoryear{{Planck Collaboration} et~al.,}{{Planck
  Collaboration} et~al.}{2016}]{Planck2016}
{Planck Collaboration} et~al., 2016, \mn@doi [\aap]
  {10.1051/0004-6361/201525830}, \href
  {http://adsabs.harvard.edu/abs/2016A%26A...594A..13P} {594, A13}

\bibitem[\protect\citeauthoryear{{Rasmussen}, {Ponman}  \&
  {Mulchaey}}{{Rasmussen} et~al.}{2006}]{Rasmussen2006}
{Rasmussen} J.,  {Ponman} T.~J.,   {Mulchaey} J.~S.,  2006, \mn@doi [\mnras]
  {10.1111/j.1365-2966.2006.10492.x}, \href
  {http://adsabs.harvard.edu/abs/2006MNRAS.370..453R} {370, 453}

\bibitem[\protect\citeauthoryear{{Rasmussen} et~al.,}{{Rasmussen}
  et~al.}{2012}]{Rasmussen2012}
{Rasmussen} J.,  et~al., 2012, \mn@doi [\apj] {10.1088/0004-637X/747/1/31},
  \href {http://adsabs.harvard.edu/abs/2012ApJ...747...31R} {747, 31}

\bibitem[\protect\citeauthoryear{{Rau} \& {Cornwell}}{{Rau} \&
  {Cornwell}}{2011}]{Rau2011}
{Rau} U.,  {Cornwell} T.~J.,  2011, \mn@doi [\aap]
  {10.1051/0004-6361/201117104}, \href
  {http://adsabs.harvard.edu/abs/2011A%26A...532A..71R} {532, A71}

\bibitem[\protect\citeauthoryear{{Reeves}, {Sadler}, {Allison}, {Koribalski},
  {Curran}  \& {Pracy}}{{Reeves} et~al.}{2015}]{Reeves2015}
{Reeves} S.~N.,  {Sadler} E.~M.,  {Allison} J.~R.,  {Koribalski} B.~S.,
  {Curran} S.~J.,   {Pracy} M.~B.,  2015, \mn@doi [\mnras]
  {10.1093/mnras/stv504}, \href
  {http://adsabs.harvard.edu/abs/2015MNRAS.450..926R} {450, 926}

\bibitem[\protect\citeauthoryear{{Rogstad}, {Lockhart}  \& {Wright}}{{Rogstad}
  et~al.}{1974}]{Rogstad1974}
{Rogstad} D.~H.,  {Lockhart} I.~A.,   {Wright} M.~C.~H.,  1974, \mn@doi [\apj]
  {10.1086/153164}, \href {http://adsabs.harvard.edu/abs/1974ApJ...193..309R}
  {193, 309}

\bibitem[\protect\citeauthoryear{{Salpeter}}{{Salpeter}}{1955}]{Salpeter1955}
{Salpeter} E.~E.,  1955, \mn@doi [\apj] {10.1086/145971}, \href
  {http://adsabs.harvard.edu/abs/1955ApJ...121..161S} {121, 161}

\bibitem[\protect\citeauthoryear{{Sancisi} \& {Allen}}{{Sancisi} \&
  {Allen}}{1979}]{Sancisi1979}
{Sancisi} R.,  {Allen} R.~J.,  1979, \aap, \href
  {http://adsabs.harvard.edu/abs/1979A%26A....74...73S} {74, 73}

\bibitem[\protect\citeauthoryear{{Sault}, {Teuben}  \& {Wright}}{{Sault}
  et~al.}{1995}]{Sault1995}
{Sault} R.~J.,  {Teuben} P.~J.,   {Wright} M.~C.~H.,  1995, in {Shaw} R.~A.,
  {Payne} H.~E.,   {Hayes} J.~J.~E.,  eds,  Astronomical Society of the Pacific
  Conference Series Vol. 77, Astronomical Data Analysis Software and Systems
  IV. p.~433 (\mn@eprint {} {astro-ph/0612759})

\bibitem[\protect\citeauthoryear{{Schiminovich} et~al.,}{{Schiminovich}
  et~al.}{2010}]{Schiminovich2010}
{Schiminovich} D.,  et~al., 2010, \mn@doi [\mnras]
  {10.1111/j.1365-2966.2010.17210.x}, \href
  {http://adsabs.harvard.edu/abs/2010MNRAS.408..919S} {408, 919}

\bibitem[\protect\citeauthoryear{{Schinckel} \& {Bock}}{{Schinckel} \&
  {Bock}}{2016}]{Schinckel2016}
{Schinckel} A.~E.~T.,  {Bock} D.~C.-J.,  2016, in Ground-based and Airborne
  Telescopes VI. p. 99062A, \mn@doi{10.1117/12.2233920}

\bibitem[\protect\citeauthoryear{Sengupta \& Balasubramanyam}{Sengupta \&
  Balasubramanyam}{2006}]{Sengupta2006}
Sengupta C.,  Balasubramanyam R.,  2006, \mn@doi [\mnras]
  {10.1111/j.1365-2966.2006.10307.x}, 369, 360

\bibitem[\protect\citeauthoryear{Serra et~al.,}{Serra et~al.}{2013}]{Serra2013}
Serra P.,  et~al., 2013, \mn@doi [Monthly Notices of the Royal Astronomical
  Society] {10.1093/mnras/sts033}, 428, 370

\bibitem[\protect\citeauthoryear{{Serra} et~al.,}{{Serra}
  et~al.}{2015a}]{Serra2015a}
{Serra} P.,  et~al., 2015a, \mn@doi [\mnras] {10.1093/mnras/stv079}, \href
  {http://adsabs.harvard.edu/abs/2015MNRAS.448.1922S} {448, 1922}

\bibitem[\protect\citeauthoryear{{Serra} et~al.,}{{Serra}
  et~al.}{2015b}]{Serra2015b}
{Serra} P.,  et~al., 2015b, \mn@doi [\mnras] {10.1093/mnras/stv1326}, \href
  {http://adsabs.harvard.edu/abs/2015MNRAS.452.2680S} {452, 2680}

\bibitem[\protect\citeauthoryear{{Sofue} \& {Rubin}}{{Sofue} \&
  {Rubin}}{2001}]{Sofue2001}
{Sofue} Y.,  {Rubin} V.,  2001, \mn@doi [\araa]
  {10.1146/annurev.astro.39.1.137}, \href
  {http://adsabs.harvard.edu/abs/2001ARA%26A..39..137S} {39, 137}

\bibitem[\protect\citeauthoryear{{Solanes}, {Manrique},
  {Garc{\'{\i}}a-G{\'o}mez}, {Gonz{\'a}lez-Casado}, {Giovanelli}  \&
  {Haynes}}{{Solanes} et~al.}{2001}]{Solanes2001}
{Solanes} J.~M.,  {Manrique} A.,  {Garc{\'{\i}}a-G{\'o}mez} C.,
  {Gonz{\'a}lez-Casado} G.,  {Giovanelli} R.,   {Haynes} M.~P.,  2001, \mn@doi
  [\apj] {10.1086/318672}, \href
  {http://adsabs.harvard.edu/abs/2001ApJ...548...97S} {548, 97}

\bibitem[\protect\citeauthoryear{{Springob} et~al.,}{{Springob}
  et~al.}{2014}]{Springob2014}
{Springob} C.~M.,  et~al., 2014, \mn@doi [\mnras] {10.1093/mnras/stu1743},
  \href {http://adsabs.harvard.edu/abs/2014MNRAS.445.2677S} {445, 2677}

\bibitem[\protect\citeauthoryear{{Taylor} et~al.,}{{Taylor}
  et~al.}{2011}]{Taylor2011}
{Taylor} E.~N.,  et~al., 2011, \mn@doi [\mnras]
  {10.1111/j.1365-2966.2011.19536.x}, \href
  {http://adsabs.harvard.edu/abs/2011MNRAS.418.1587T} {418, 1587}

\bibitem[\protect\citeauthoryear{{Thilker} et~al.,}{{Thilker}
  et~al.}{2007}]{Thilker2007}
{Thilker} D.~A.,  et~al., 2007, \mn@doi [\apjs] {10.1086/523853}, \href
  {http://adsabs.harvard.edu/abs/2007ApJS..173..538T} {173, 538}

\bibitem[\protect\citeauthoryear{{Tully}}{{Tully}}{1987}]{Tully1987}
{Tully} R.~B.,  1987, \mn@doi [\apj] {10.1086/165629}, \href
  {http://adsabs.harvard.edu/abs/1987ApJ...321..280T} {321, 280}

\bibitem[\protect\citeauthoryear{{Vogelaar} \& {Terlouw}}{{Vogelaar} \&
  {Terlouw}}{2001}]{Vogelaar2001}
{Vogelaar} M.~G.~R.,  {Terlouw} J.~P.,  2001, in {Harnden} Jr. F.~R.,
  {Primini} F.~A.,   {Payne} H.~E.,  eds,  Astronomical Society of the Pacific
  Conference Series Vol. 238, Astronomical Data Analysis Software and Systems
  X. p.~358

\bibitem[\protect\citeauthoryear{Westmeier, Braun  \& Koribalski}{Westmeier
  et~al.}{2011}]{Westmeier2011}
Westmeier T.,  Braun R.,   Koribalski B.~S.,  2011, \mn@doi [\mnras]
  {doi:10.1111/j.1365-2966.2010.17596.x}, 410, 2217

\bibitem[\protect\citeauthoryear{Westmeier, Koribalski  \& Braun}{Westmeier
  et~al.}{2013}]{Westmeier2013}
Westmeier T.,  Koribalski B.~S.,   Braun R.,  2013, \mn@doi [\mnras]
  {10.1093/mnras/stt1271}, 434, 3511

\bibitem[\protect\citeauthoryear{Wevers, Appleton, Davies  \& Hart}{Wevers
  et~al.}{1984}]{Wevers1984}
Wevers B. M. H.~R.,  Appleton P.~N.,  Davies R.~D.,   Hart L.,  1984, a{\&}a,
  140, 125

\bibitem[\protect\citeauthoryear{{Wilson} et~al.,}{{Wilson}
  et~al.}{2011}]{Wilson2011}
{Wilson} W.~E.,  et~al., 2011, \mn@doi [\mnras]
  {10.1111/j.1365-2966.2011.19054.x}, \href
  {http://adsabs.harvard.edu/abs/2011MNRAS.416..832W} {416, 832}

\bibitem[\protect\citeauthoryear{{Wong}, {Meurer}, {Zheng}, {Heckman},
  {Thilker}  \& {Zwaan}}{{Wong} et~al.}{2016}]{Wong2016}
{Wong} O.~I.,  {Meurer} G.~R.,  {Zheng} Z.,  {Heckman} T.~M.,  {Thilker} D.~A.,
    {Zwaan} M.~A.,  2016, \mn@doi [\mnras] {10.1093/mnras/stw993}, \href
  {http://adsabs.harvard.edu/abs/2016MNRAS.460.1106W} {460, 1106}

\bibitem[\protect\citeauthoryear{{Wright} et~al.,}{{Wright}
  et~al.}{2016}]{Wright2016}
{Wright} A.~H.,  et~al., 2016, \mn@doi [\mnras] {10.1093/mnras/stw832}, \href
  {http://adsabs.harvard.edu/abs/2016MNRAS.460..765W} {460, 765}

\bibitem[\protect\citeauthoryear{{Zabludoff} \& {Mulchaey}}{{Zabludoff} \&
  {Mulchaey}}{1998}]{Zabludoff1998}
{Zabludoff} A.~I.,  {Mulchaey} J.~S.,  1998, \mn@doi [\apj] {10.1086/305355},
  \href {http://adsabs.harvard.edu/abs/1998ApJ...496...39Z} {496, 39}

\bibitem[\protect\citeauthoryear{{Zwaan}, {Meyer}, {Staveley-Smith}  \&
  {Webster}}{{Zwaan} et~al.}{2005}]{Zwaan2005}
{Zwaan} M.~A.,  {Meyer} M.~J.,  {Staveley-Smith} L.,   {Webster} R.~L.,  2005,
  \mn@doi [\mnras] {10.1111/j.1745-3933.2005.00029.x}, \href
  {http://adsabs.harvard.edu/abs/2005MNRAS.359L..30Z} {359, L30}

\bibitem[\protect\citeauthoryear{{de Blok}, {McGaugh}  \& {Rubin}}{{de Blok}
  et~al.}{2001}]{deblok2001}
{de Blok} W.~J.~G.,  {McGaugh} S.~S.,   {Rubin} V.~C.,  2001, \mn@doi [\aj]
  {10.1086/323450}, \href {http://adsabs.harvard.edu/abs/2001AJ....122.2396D}
  {122, 2396}

\bibitem[\protect\citeauthoryear{{de Blok}, {Walter}, {Brinks}, {Trachternach},
  {Oh}  \& {Kennicutt}}{{de Blok} et~al.}{2008}]{deblok2008}
{de Blok} W.~J.~G.,  {Walter} F.,  {Brinks} E.,  {Trachternach} C.,  {Oh}
  S.-H.,   {Kennicutt} Jr. R.~C.,  2008, \mn@doi [\aj]
  {10.1088/0004-6256/136/6/2648}, \href
  {http://adsabs.harvard.edu/abs/2008AJ....136.2648D} {136, 2648}

\bibitem[\protect\citeauthoryear{{de Blok} et~al.,}{{de Blok}
  et~al.}{2017}]{deblok2017}
{de Blok} W.~J.~G.,  et~al., 2017, preprint, \href
  {http://adsabs.harvard.edu/abs/2017arXiv170908458D} {} (\mn@eprint {arXiv}
  {1709.08458})

\bibitem[\protect\citeauthoryear{{de Vaucouleurs}, {de Vaucouleurs}, {Corwin},
  {Buta}, {Paturel}  \& {Fouqu{\'e}}}{{de Vaucouleurs}
  et~al.}{1991}]{devaucouleurs1991}
{de Vaucouleurs} G.,  {de Vaucouleurs} A.,  {Corwin} Jr. H.~G.,  {Buta} R.~J.,
  {Paturel} G.,   {Fouqu{\'e}} P.,  1991, {Third Reference Catalogue of Bright
  Galaxies. Volume I: Explanations and references. Volume II: Data for galaxies
  between 0$^{h}$ and 12$^{h}$. Volume III: Data for galaxies between 12$^{h}$
  and 24$^{h}$.}

\bibitem[\protect\citeauthoryear{{van der Hulst}, {Terlouw}, {Begeman},
  {Zwitser}  \& {Roelfsema}}{{van der Hulst} et~al.}{1992}]{vanderHulst1992}
{van der Hulst} J.~M.,  {Terlouw} J.~P.,  {Begeman} K.~G.,  {Zwitser} W.,
  {Roelfsema} P.~R.,  1992, in {Worrall} D.~M.,  {Biemesderfer} C.,   {Barnes}
  J.,  eds,  Astronomical Society of the Pacific Conference Series Vol. 25,
  Astronomical Data Analysis Software and Systems I. p.~131

\bibitem[\protect\citeauthoryear{{van der Kruit} \& {Searle}}{{van der Kruit}
  \& {Searle}}{1981a}]{Kruit1981b}
{van der Kruit} P.~C.,  {Searle} L.,  1981a, \aap, \href
  {http://adsabs.harvard.edu/abs/1981A%26A....95..105V} {95, 105}

\bibitem[\protect\citeauthoryear{{van der Kruit} \& {Searle}}{{van der Kruit}
  \& {Searle}}{1981b}]{Kruit1981a}
{van der Kruit} P.~C.,  {Searle} L.,  1981b, \aap, \href
  {http://adsabs.harvard.edu/abs/1981A%26A....95..116V} {95, 116}

\makeatother
\end{thebibliography}

%\begin{thebibliography}{99}
%\end{thebibliography}

%%%%%%%%%%%%%%%%%%%%%%%%%%%%%%%%%%%%%%%%%%%%%%%%%%

%%%%%%%%%%%%%%%%% APPENDICES %%%%%%%%%%%%%%%%%%%%%

\appendix

%%%%%%%%%%%%%%%%%%%%%%%%%%%%%%%%%%%%%%%%%%%%%%%%%%

% Don't change these lines
\bsp	% typesetting comment
\label{lastpage}
\end{document}